\documentclass[nofootinbib,amsfonts,floatfix,superscriptaddress,longbibliography,twocolumn]{revtex4-2}

\usepackage{mathbbol}
\usepackage{amsmath,amssymb,bm,amsthm}
\usepackage{epstopdf}
\usepackage{braket}
\usepackage{soul}

\usepackage[utf8x]{inputenc}
\usepackage[usenames,dvipsnames, pdftex]{xcolor}
\usepackage[colorlinks, linkcolor=blue,citecolor=blue, urlcolor=blue]{hyperref}
\usepackage{graphicx}
\usepackage{subfigure}
\definecolor{magenta}{rgb}{1.0, 0.0, 0.56}

\begin{document}

\title{Interacting bosons in a triple well: Preface of many-body quantum chaos}

\author{ Karin Wittmann W.}
\address{Instituto de F\'{i}sica da UFRGS, Porto Alegre, RS, Brazil}
\author{E. R. Castro}
\address{Instituto de F\'{i}sica da UFRGS, Porto Alegre, RS, Brazil}
\author{Angela Foerster}
\address{Instituto de F\'{i}sica da UFRGS, Porto Alegre, RS, Brazil}
\author{Lea F. Santos}
\affiliation{Department of Physics, Yeshiva University, New York, New York 10016, USA}

\begin{abstract}
Systems of interacting bosons in triple-well potentials are of significant theoretical and experimental interest. They are explored in contexts that range from quantum phase transitions and quantum dynamics to semiclassical analysis.  Here, we systematically investigate the onset of quantum chaos in a triple-well model that moves away from integrability as its potential gets tilted. Even in its deepest chaotic regime, the system presents features reminiscent of integrability. Our studies are based on level spacing distribution and spectral form factor, structure of the eigenstates,  diagonal and off-diagonal elements of observables in relationship to the eigenstate thermalization hypothesis. With only three sites, the system's eigenstates are at the brink of becoming fully chaotic and thus show distributions other than Gaussian, which resonates with the results for the observables.
\end{abstract}
\maketitle

\section{Introduction}
The interest in many-body quantum chaos has grown significantly in recent years due to its close connection with thermalization~\cite{Zelevinsky1996,Borgonovi2016,Dalessio2016}, scrambling of quantum information~\cite{Maldacena2016JHEP}, and the fact that many-body quantum systems can now be studied experimentally in a controllable way with a variety of experimental set-ups, from cold atoms and ion traps to superconducting devices and nuclear magnetic resonance. In studies of the subject, the focus is usually on interacting lattice systems with many sites and many particles, where the Hilbert space grows exponentially with the system size. Here, instead, we investigate the onset of quantum chaos in a system that has only three wells, but where the number $N$ of particles is large. The Hilbert space grows quadratically with $N$, and as this number increases, the system is brought closer to the classical limit. 

Fascinating phenomena are explored with systems of interacting atoms in triple-well potentials, such as transistor-like behaviors~\cite{Zhang2015, Caliga2016,Marchukov2016}, entanglement generation~\cite{Viscondi2010,Tonel2020}, coherent population transfer~\cite{Bradly2012,Xiong2013,Zhou2013, Olsen2014},  fragmentation~\cite{Gallemi2013, Gallemi2015}, quantum-classical correspondence~\cite{Nemoto2000,Liu2007,Castro2021}, quantum chaos~\cite{March2018,Richaud2018,Bera2019,Ray2020,rautenberg2020classical,Nakerst2021}, and caustics~\cite{kirkby2021caustics}, among others~\cite{Foerster_2007,Buonsante2009,Lahaye2010,Streltsov2011,Viscondi2011,Cao2011,Guo2014,Koutentakis2017,Guo2018,Dutta2019,McCormack2020}.
One of the most popular models in this context  is the three-well Bose-Hubbard model with short-range interactions and local hopping terms~\cite{Fisher1989, Bloch2005}. This system, with three or more wells, is in general not integrable \cite{Choy1982,Kolovsky2004, Oelkers2007,kollath2010statistical,Nakerst2021}. Integrability is achieved with two or with an infinite number of wells.

A bosonic triple-well model with an integrable limit was introduced in Ref.~\cite{Ymai2017} and explored for switching devices~\cite{Wittmann2018,foot4}. This model is a member of a family of quantum integrable multi-well tunneling systems that have the two-site Bose-Hubbard model~\cite{Leggett2001, Tonel2005,Links2006} as a leading constituent. Integrability requires the presence of long-range  couplings, which is in fact a physical condition for ultracold dipolar bosons with large dipole moment, such as chromium, erbium, or dysprosium.  Dipolar cold atoms provide a rich platform for the study of mesoscopic quantum superpositions~\cite{Lahaye2010}, macroscopic cat states~\cite{DellAnna2013}, quantum droplets~\cite{Ferrier2016}, and supersolid states~\cite{Chomaz2019}. 
By tilting the potential, the bosonic triple-well model introduced in~\cite{Ymai2017} becomes chaotic. We provide a systematic study of this transition based not only on spectral correlations, but also on the structure of the eigenstates and its consequences to the eigenstate expectation values and the distributions of the off-diagonal elements of the number operator of each well. 

Contrary to interacting multi-well systems with many particles, in the  triple-well model, the range of the values of the integrability breaking parameter that leads to chaos does not increase as the Hilbert space grows, and  the eigenstates do not reach a high degree of chaoticity. Even in the middle of the chaotic region, the eigenstates still present some level of correlation and part of their components lie in the non-chaotic regions of the spectrum. The chaotic features of the triple-well model cannot be enhanced by increasing  the number of bosons. 

In non-driven systems, three wells constitute the turning point for the onset of many-body quantum chaos. The transition from integrability to chaos does take place, but with some reminiscence of integrability. 

The paper is organized as follows. The model is described in Sec.~II. The analysis of the spectrum and level repulsion are presented in Sec.~III. The core of the work is the detailed study of the structure of the eigenstates in Sec.~IV and its consequence to the diagonal and off-diagonal elements of the number operators in Sec.~V. Conclusions are given in Sec.~VI.

\section{Model}

The quantum system that we study consists of $N$ bosons in an aligned triple-well potential described by the  following Hamiltonian, 

\begin{align}
\label{QH}
	\hat{H} =& \frac{U}{N}\left(\hat{N}_1^2-\hat{N}_2^2+\hat{N}_3^2\right)
	+ \epsilon\left(\hat{N}_3-\hat{N}_1\right) \nonumber \\
	&+ \frac{2 U}{N} \left( - \hat{N}_1 \hat{N}_2 - \hat{N}_2 \hat{N}_3 + \hat{N}_1 \hat{N}_3 \right)
	\nonumber \\
	&+\frac{J}{\sqrt{2}}\left(\hat{a}_1^\dagger \hat{a}_2 + \hat{a}_2^\dagger \hat{a}_1\right)+\frac{J}{\sqrt{2}}\left(\hat{a}_2^\dagger \hat{a}_3 + \hat{a}_3^\dagger \hat{a}_2\right),
\end{align}
where $\hat{N}_i=\hat{a}_i^\dagger \hat{a}_i$ is the number operator of the well $i$, $\hat{a}_i$ ($\hat{a}_i^\dagger$) is the annihilation (creation) operator, $U$ is the onsite interaction strength and also the strength of the interactions between wells,  $J$ is the tunneling amplitude between wells, and $\epsilon$ is the amplitude of the tilt between the wells. We consider repulsive interaction, $U\geq0$.

Hamiltonian (\ref{QH}) conserves  the total number of bosons, $N=N_1+N_2+N_3$, and when $\epsilon=0$, it commutes with the parity operator. The matrix has dimension $D=(N+2)!/(2!N!)$. Our studies of the structure of the eigenstates are done in the Fock basis, $|n\rangle = |N_1, N_2, N_3 \rangle$. We denote the eigenstates and eigenvalues of $\hat{H}$ by $|\alpha \rangle$ and $E_{\alpha}$. 

\begin{figure}[h]
	\centering
		\includegraphics[width=1.\linewidth]{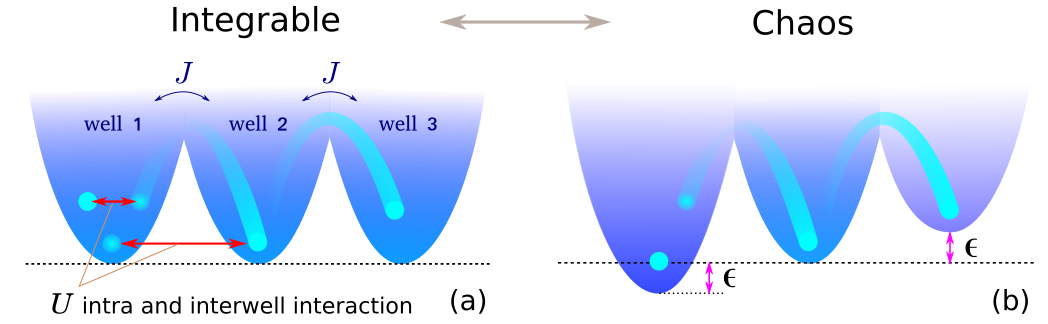}
	\caption{Schematic representation of the three-well system described by Eq.~(\ref{QH})  for both the integrable (a) and the non-integrable (b) regime. The red arrows indicate the intrawell and interwell interaction strength $U$, the black arrows indicate the tunneling amplitude $J$ between adjacent wells, and $\epsilon$ represents the tilt of the potentials of wells 1 and 3 with respect to well 2. }
		\label{fig01}
	\end{figure}

A schematic representation of our model is shown in Fig.~\ref{fig01}. When $\epsilon=0$ [Fig.~\ref{fig01}~(a)], the model is integrable and solvable with the algebraic Bethe ansatz~\cite{Ymai2017}. At this point, in addition to energy and the total number of particles, our three-degree-of-freedom model has a third independent conserved quantity, $Q = J_1^2 N_3 + J_3^2 N_1 - J_1 J_3(a_1^\dagger a_3+a_3^\dagger a_1)$ \cite{Ymai2017, Wittmann2018}. The system becomes nonintegrable [Fig.~\ref{fig01}~(b)] when the tilt is included. As discussed in Sec.~\ref{LevelSta}, the model shows signatures of quantum chaos when the tilt amplitude is of the order of the hopping and interaction strengths, $\epsilon \sim J, U$. 

In the absence of the potential tilt and of the interaction between wells, our model coincides with the bare Bose-Hubbard model with 3 sites. Signatures of quantum chaos were studied in this model with 5 sites and 5 bosons~\cite{Kolovsky2004} and more recently for only 3 sites and $N \gg3$ \cite{Nakerst2021}. The latter case and also the extended triple-well Bose-Hubbard model with dipolar interaction~\cite{Lahaye2010} exhibit properties similar to those of our system in the chaotic domain. Comparisons between the three models are presented in the appendix~\ref{appA}.

\subsection{Parameters and density of states}

\begin{figure}[h]
	\centering
	\includegraphics[width=0.49\linewidth]{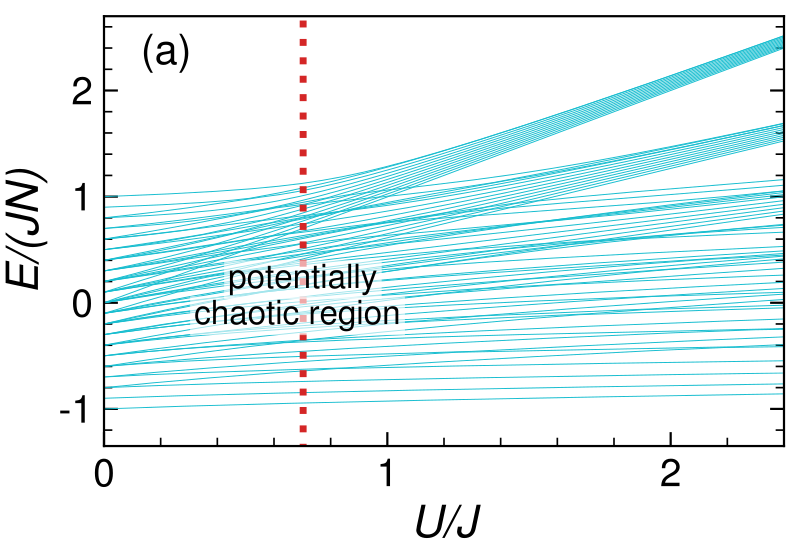}	
	\includegraphics[width=0.49\linewidth]{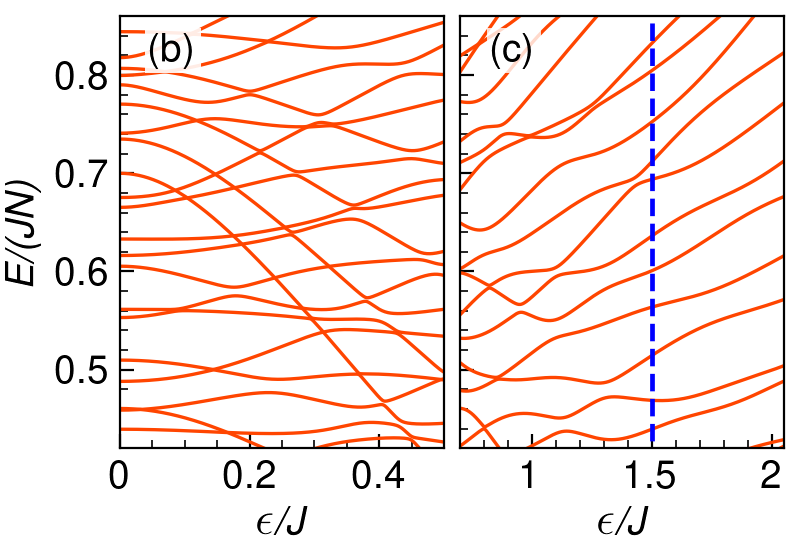}
	\caption{Normalized eigenvalues as a function of the interaction strength, $U/J$, for $\epsilon=0$ (a) and as a function of the tilt amplitude, $\epsilon/J$, for $U/J=0.7$ (b)-(c). The vertical line in Fig.~\ref{fig02}~(a) marks the value $U/J=0.7$, which is used in all of our subsequent studies. In Fig.~\ref{fig02}~(c), the vertical line marks the value $\epsilon/J=1.5$ used in our studies of the chaotic regime. In all panels N=10.}
		\label{fig02}
	\end{figure}

In our numerical analysis, we fix $J=1$, $U/J=0.7$, and vary $\epsilon$ for different numbers of particles.  The choice of $U$ is justified with Fig.~\ref{fig02}~(a), where we show the eigenvalues as a function of the interaction strength for $\epsilon=0$.  When $U=0$, there is only hopping and the model is trivially solved. This is usually referred to as Rabi regime~\cite{Wittmann2018} in analogy with the double-well model~\cite{Leggett2001, Tonel_2005b}.  As the interaction strength increases and becomes larger than the hopping amplitude, $U/J>1$, energy bands are formed. The extreme scenario of $U \gg J$ is the Fock regime, where the eigenstates approach the Fock states, and the model is again trivially solved. The region where we can expect chaos to develop is therefore for $0<U/J<1$, which explains the choice $U/J=0.7$ indicated with the red dotted vertical line in Fig.~\ref{fig02}~(a).

In chaotic systems, the eigenvalues are correlated and avoid each other~\cite{HaakeBook,Guhr1998}, while in integrable models (apart from the picket-fence scenario~\cite{Berry1977,Pandey1991,Chirikov1995}), the levels can cross. This difference is clearly seen in Fig.~\ref{fig02}~(b) and Fig.~\ref{fig02}~(c), where we fix $U/J=0.7$ and vary $\epsilon/J$. Level crossing happens when $0\leq \epsilon/J<1$ [Fig.~\ref{fig02}~(b)], but is avoided when $\epsilon/J \sim 1$ [Fig.~\ref{fig02}~(c)]. This latter panel exhibits the ``spaghetti structure'' typical of repulsive energy levels.

\begin{figure}[h]
	\centering
	\includegraphics[width=1.\linewidth]{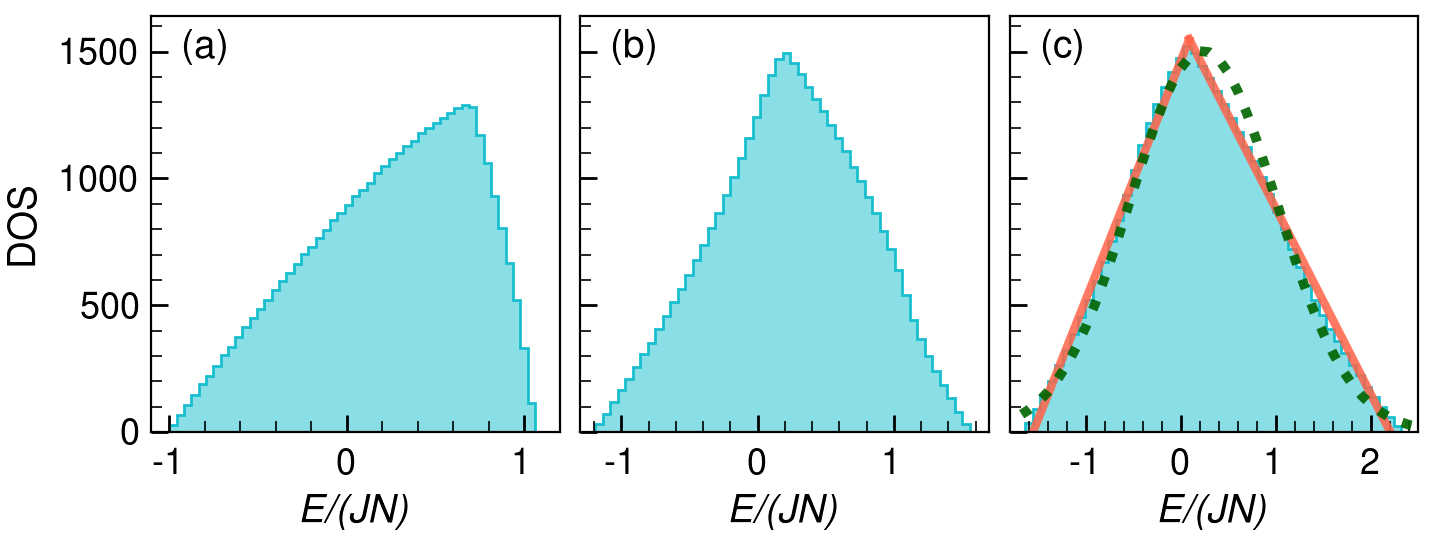}
	\caption{Density of states for $N=270$, $U/J=0.7$, and $\epsilon/J=0$ (a), $0.7$ (b) and $1.5$ (c). The solid line in (c) is a linear fitting for the left and right sides of the distribution, while the dotted line is a Gaussian distribution fitting. 
	}
		\label{fig03}
	\end{figure}

In Fig.~\ref{fig03}, we compare the density of states (DOS),
\begin{equation}
\nu (E) = \sum_{\alpha=1}^{D} \delta(E - E_{\alpha}),
\end{equation}
of the model~(\ref{QH}) for three values of the tilt, $\epsilon/J =0$, $0.7$, and $1.5$. In realistic interacting many-body quantum systems with  many degrees of freedom, such as spin models with many excitations~\cite{Santos2012PRE} or Bose-Hubbard models with many particles and many sites~\cite{Kolovsky2004,Pausch2021}, the DOS is typically Gaussian~\cite{French1970,Brody1981}, which can be explained using the central limit theorem. This contrasts with our model [Figs.~\ref{fig03}~(a)-(c)], which has few degrees of freedom. 

Systems with few-degrees of freedom, such as the Dicke model~\cite{Bastarrachea2014}, spin-1/2 models with less than 4 excitations~\cite{Schiulaz2018}, and  multi-well Kronig-Penney-like systems with few particles~\cite{Fogarty2021}, often present shapes other than Gaussian.  We see in the appendix~\ref{appA} that the bare triple-well Bose-Hubbard model and the extended  triple-well Bose-Hubbard model show distributions that, similarly to our model in Fig.~\ref{fig03}~(c), are not yet Gaussian, but get close to it. The DOS for the extended  Bose-Hubbard model and for our model are comparable, since both have long-range couplings.

\section{Spectral correlations}

To quantify the degree of correlations between the eigenvalues, we study the level spacing distribution and the spectral form factor. We show that for $U/J\sim 0.7$, as $\epsilon$ in Eq.~(\ref{QH}) increases from zero, our triple-well model leaves the integrable point ($\epsilon=0$) and moves towards the chaotic domain.

\subsection{Level spacing distribution}
\label{LevelSta}

The transition to quantum chaos can be verified with the distribution $P(s)$ of the spacings $s$ between nearest unfolded energy levels. For chaotic systems with real and symmetric Hamiltonian matrices, as in Eq.~(\ref{QH}), $P(s)$ follows the Wigner  surmise~\cite{MehtaBook,Guhr1998}, $P_{\rm W}(s) = (\pi s/2) \exp \left( -\pi s^2/4 \right)$, as obtained also for the eigenvalues of full random matrices from a Gaussian orthogonal ensemble (GOE). This distribution indicates that the eigenvalues are correlated and repel each other, that is, $P(s =0) = 0$.  In integrable models, the level spacing distribution is Poissonian, $P_{\rm P}(s) = e^{-s}$, since the energy levels are uncorrelated~\cite{foot01}.

The analysis of the level spacing distribution requires unfolding the eigenvalues and separating them by symmetry sectors. The unfolding procedure corresponds to rescaling the eigenvalues, so that the local density of states of the rescaled energies is 1.  The separation by subspaces is necessary, because eigenvalues from different symmetry sectors have no reason to be correlated.

\begin{figure}[h]
	\centering
		\includegraphics[width=1.\linewidth]{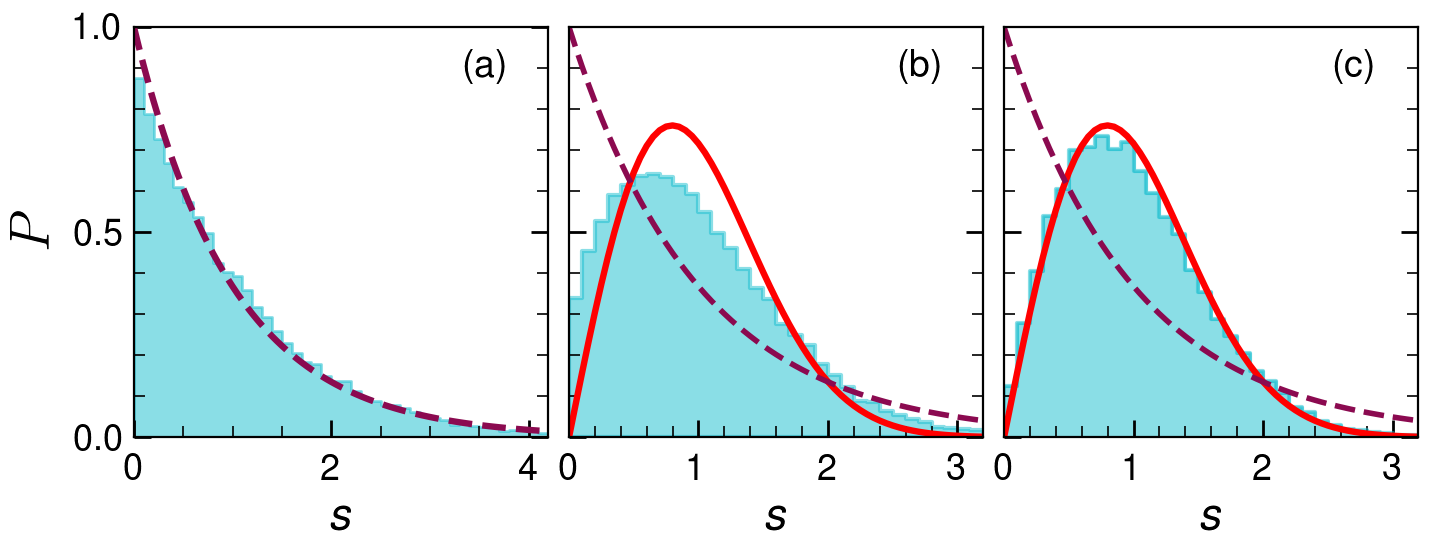}\\
		\includegraphics[width=0.8\linewidth]{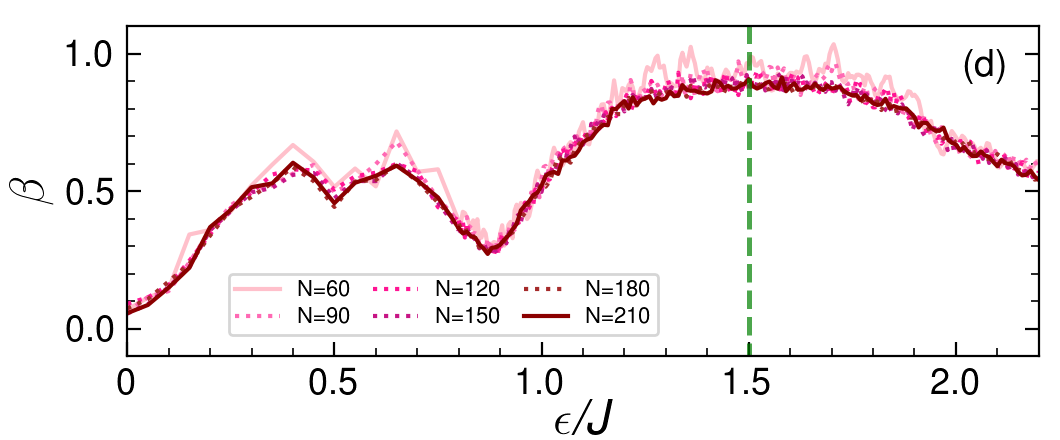}
	\caption{Level spacing distribution for $N=270$ and $\epsilon/J=0$ (a),  $\epsilon/J=0.7$ (b), and $\epsilon/J=1.5$ (c); and chaos indicator $\beta$ as a function of the tilt amplitude for various $N$'s (d). In (a)-(c): The dashed (solid) line represents the Poissonian (Wigner) distribution.  In (d): The green vertical line at $\epsilon/J=1.5$ marks where $\beta$ gets the closest to 1, indicating the Wigner distribution.}
		\label{fig04}
	\end{figure}

In Figs.~\ref{fig04}~(a)-(c), we illustrate $P(s)$ for $\epsilon/J =0$, $0.7$, and $1.5$, respectively. The Poissonian distribution is obtained for the integrable point $\epsilon=0$ in Fig.~\ref{fig04}~(a), and the Wigner shape is seen for $\epsilon/J=1.5$ in Fig.~\ref{fig04}~(c), as we had anticipated from the ``spaghetti structure'' in Fig.~\ref{fig02}~(c). An intermediate picture emerges for $\epsilon/J=0.7$ in Fig.~\ref{fig04}~(b).

The proximity of the level spacing distribution to the Poissonian or the Wigner distribution can be quantified with the chaos indicator $\beta$, which is obtained by fitting $P(s)$ with the Brody distribution~\cite{Brody1981} (see also ~\cite{Izrailev1990}),
\begin{equation}
P_\beta (s) = (\beta+1) b s^{\beta} \exp(-b s^{\beta+1}), \hspace{0.4 cm} 
b =\left[ \Gamma \left( \frac{\beta+2}{\beta+1} \right) \right]^{\beta+1}
\label{Eq:beta}
\end{equation}
For chaotic systems, $\beta \sim 1$ and for a Poissonian distribution, $\beta \sim 0$. 

In Fig.~\ref{fig04}~(d), we show $\beta$ as a function of $\epsilon/J$ for $N=60$, $90$, $\ldots 210$. As evident from the figure, a high degree of chaos happens for $\epsilon/J \in [1.3,1.7]$. Notice that this range of values does not grow as $N$ increases, which contrasts with interacting many-body quantum systems with many sites~\cite{Santos2010PRE,RigolSantos2010,Santos2020}, where studies of chaos indicators for different system sizes suggest that in the thermodynamic limit, an infinitesimal integrability breaking term may be enough to bring those systems to the chaotic domain. In addition and also contrary to the results for systems with many sites~\cite{RigolSantos2010,Santos2020}, larger values of $N$ do not take $\beta$ closer to 1. The only effect that an increased value of $N$ appears to have for the triple-well model is to reduce the fluctuations in the values of $\beta$ for nearby $\epsilon$'s, which concurs with improved statistics.

\subsection{Spectral form factor}

The level spacing distribution detects only short-range correlations. To get a better idea of the degree of spectral correlations, one may resort to other indicators of quantum chaos, such as the spectral form factor,
\begin{equation}
S_{FF} (t) = \left\langle \left| \sum_{\alpha=1}^D f(E_\alpha) e^{-i E_{\alpha}  t} \right|^2 \right\rangle, 
\label{Eq:SFF}
\end{equation}
which captures both short- and long-range correlations. The spectral form factor is used to study level statistics in the time domain. When the eigenvalues correlated as in random matrices, $S_{FF}(t)$ develops the so-called correlation hole~\cite{Leviandier1986,Guhr1990,Wilkie1991,Alhassid1992,Gorin2002,Torres2017Philo}, which we further discuss below Eq.~(11). The spectral form factor is advantageous over the direct analysis of the eigenvalues, because it does not require unfolding the spectrum or separating the eigenvalues by symmetry sectors~\cite{Cruz2020,Santos2020}, although averages, indicated by $\langle . \rangle$ in Eq.~(\ref{Eq:SFF}), are needed, since this quantity is non-self-averaging~\cite{Prange1997,Schiulaz2020}. 

A filter function $f(E_{\alpha})$, as used in Eq.~(\ref{Eq:SFF}), is often added to the spectral form factor~\cite{WinerARXIV}. When $f(E_{\alpha})$ coincides with the components of an initial state projected in the energy eigenbasis, the spectral form factor becomes the survival probability~\cite{Schiulaz2020}.  In our analysis, we choose~\cite{Lerma2019} 
\begin{equation}
f(E_\alpha) = \frac{r_\alpha g(E_\alpha)}{\sum_\beta r_\beta g(E_\beta)} ,
\end{equation}
where $r_\alpha$ are random numbers from a uniform distribution in the interval $[0,1]$, the function $g(E)=\rho(E)/\nu(E)$, and $\rho(E)$ is a chosen energy profile, which, in our case, is a rectangular function,
\begin{eqnarray}
\rho(E)&=& \left\{\begin{array}{ll}\dfrac{1}{2\sigma}&  {\hbox  { for $E \in [E_c-\sigma,E_c+\sigma]$}} 
\\
\\
0&  {\hbox   { otherwise}} , \end{array} \right.
\label{Eq:rect}
\end{eqnarray}
of width $\sigma$, centered at the energy $E_c$, and with bounds at $E_{min}= E_c - \sigma$ and $E_{max}= E_c + \sigma $. The division of $\rho(E)$ by $\nu(E)$ is done using the linear fits for the DOS in Fig.~\ref{fig03}~(c). This procedure compensates for variations in the density of states and ensures the rectangular shape of the filter function~\cite{Lerma2019}. As it will become clear in Sec.~\ref{Sec:Eigenstates}, the region where the eigenstates are mostly chaotic happens for $E/(JN) \in [-0.2,1]$. For this reason, we choose $E_c/(JN)=0.5$ and $\sigma/(JN)=0.35$. 

\begin{figure}[h]
	\centering
	\includegraphics[width=1\linewidth,trim = {0 4.6cm 0 0}, clip]{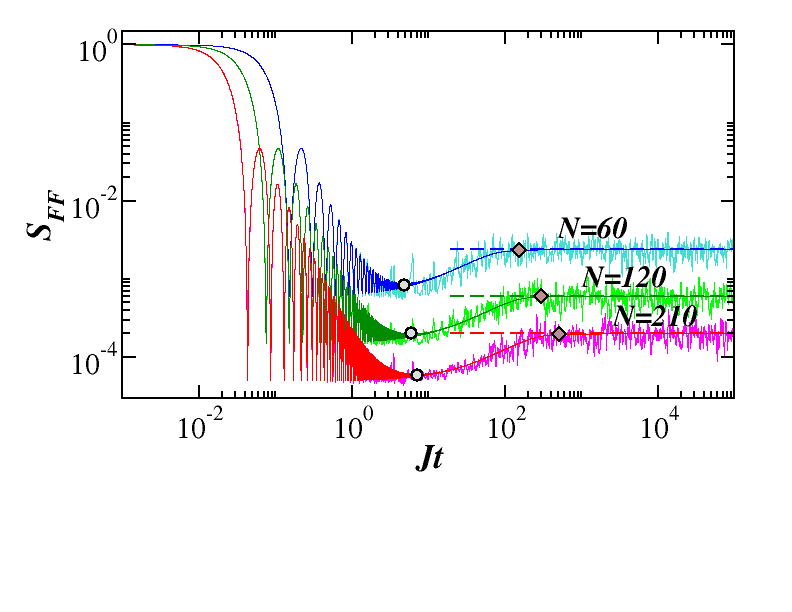}
	\caption{Spectral form factor for three values of $N$. The lines with fluctuations, which have lighter colors, represent numerical results; the thin smooth lines give the analytical expression in Eq.~(\ref{Eq:analytics}); and the dashed horizontal lines indicate the saturation point $\langle \overline{S_{FF}}  \rangle$ in Eq.~(\ref{Eq:sat}).
	The symbols mark the time to reach the minimum of the correlation (circle) and the saturation time (diamond). For the numerical results: 
	 Averages over 500 random realizations and also running averages.}
		\label{fig05}
	\end{figure}

In Fig.~\ref{fig05}, we show $S_{FF}(t)$ in the chaotic domain averaged over various realizations of the random numbers $r_{\alpha}$ and taking into account also a moving time average starting at $t\sim5/J$, where the fluctuations are large.  The numerical results  are presented together with the analytical expression obtained following Refs.~\cite{Torres2018,Schiulaz2019,Lerma2019},
\begin{equation}
S_{FF}^{analyt} (t) =\frac{1- \langle \overline{S_{FF}}  \rangle }{\eta-1}
\left[\eta \frac{\sin^2(\sigma t)}{(\sigma t)^2}     -     b_{2}\left(\frac{t}{2\pi \nu_c }\right)\right] 
+ \langle \overline{S_{FF}}  \rangle,
\label{Eq:analytics}
\end{equation}   
where 
\[
\eta = \frac{\langle r_{\alpha}^2 \rangle }{\langle r_{\alpha} \rangle^2 \langle \overline{S_{FF}}  \rangle} = \frac{4}{3 \langle \overline{S_{FF}}  \rangle}
\]
is the effective dimension associated with the chosen filter function, and 
\begin{equation}
\nu_c = \frac{\eta}{2 \sigma}
\end{equation}
is the density of states at $E=E_c$, or equivalently, the inverse mean level spacing  probed by the chosen energy profile~\cite{Lerma2019}. 

The first term in the square brackets of Eq.~(\ref{Eq:analytics}) describes the behavior of $S_{FF}(t)$ at short times. It is obtained by writing Eq.~(\ref{Eq:SFF}) as an integral,
\begin{equation}
S_{FF} (t) = \left\langle \left| \int_{E_{min}}^{E_{max}} \rho_0(E) e^{-i E t} dE \right|^2 \right\rangle, 
\label{Eq:SFFc}
\end{equation}
and substituting the energy distribution, 
\begin{equation}
\rho_0 (E) = \sum_{\alpha=1}^D f(E_\alpha) \delta(E- E_{\alpha}  ) ,
\end{equation}
with the smoothed energy profile $\rho(E)$ from Eq.~(\ref{Eq:rect}). The Fourier transform in Eq.~(\ref{Eq:SFFc}) gives $ \dfrac{\sin^2(\sigma t)}{(\sigma t)^2} $. This function leads to a power-law decay with exponent $2$ due to the bounds of the filter function~\cite{Khalfin1958,Tavora2016,Yang2020}.

The effects of the spectral correlations get manifested at larger times, when the discreteness of the spectrum is resolved and the correlations are then detected. This results in the dip in Fig.~\ref{fig05} below the horizontal dashed line that represents the infinite-time average
\begin{equation}
\langle \overline{S_{FF}}  \rangle= \sum_\alpha |f(E_\alpha)|^2.
\label{Eq:sat}
\end{equation}
This dip is known as correlation hole~\cite{Leviandier1986,Guhr1990,Wilkie1991,Alhassid1992,Gorin2002,Torres2017Philo} and it does not exist in models that present a Poissonian level spacing distribution.  In the case of GOE full random matrices, the dip is described by the two-level form factor~\cite{MehtaBook},
\begin{equation}
b_2(\bar{t}) =\begin{cases}
1-2\bar{t} + \bar{t} \ln(2 \bar{t}+1) & \bar{t}\leq1 \\
\bar{t} \ln \left( \dfrac{2 \bar{t}+1}{2 \bar{t} -1} \right) -1 & \bar{t}>1
\end{cases}.
\label{Eq:hole}
\end{equation} 
This function describes very well our numerical results and confirms the chaoticity of our triple-well model.

By comparing the results for different numbers of bosons in Fig.~\ref{fig05}, it is clear that the time to reach the minimum of the correlation hole and the time to reach saturation increase with $N$. Analytical expressions for these times are given in the appendix~\ref{appFORM}. They are much shorter than those obtained for interacting many-body quantum systems with many sites~\cite{Schiulaz2019}.

\section{Eigenstates}

\label{Sec:Eigenstates}

In chaotic quantum systems, the eigenvalues are correlated and the eigenstates are uncorrelated.
In this section, we analyze the transition to quantum chaos through the changes in the structure of the eigenstates. As $\epsilon$ increases from zero and the system moves from the integrable to the chaotic domain, we expect the eigenstates away from the edges of the spectrum to become closer to the eigenstates of GOE full random matrices~\cite{Zelevinsky1996,Borgonovi2016}. The GOE eigenstates are random vectors with components that correspond to independent Gaussian random numbers satisfying the normalization condition. In realistic many-body quantum systems, a fraction of the components of the chaotic eigenstates can be nearly zero, but the nonzero components  follow a Gaussian distribution~\cite{Santos2012PRE}. To detect the onset of these chaotic eigenstates, one can employ measures of delocalization~\cite{Izrailev1990,Torres2016Entropy} and fractality~\cite{Pausch2021}, and analyze the distributions of the components of the eigenstates. These methods are, of course, attached to a basis choice. We use the Fock basis, $|n\rangle$, which are the eigenstates of the number operators studied in Sec.~V.

Our results show that even the most delocalized eigenstates of our triple-well model are not fully chaotic. A similar conclusion was reached for the triple-well Bose-Hubbard model in Ref.~\cite{Nakerst2021}. The anomalous scaling of the eigenstate-to-eigenstate fluctuations of expectation values of local observables with  the Hilbert space found in that work can be attributed to eigenstates that are not fully chaotic.

\subsection{Delocalization Measures}

\begin{figure}[h]
	\centering
	\includegraphics[width=1.\linewidth]{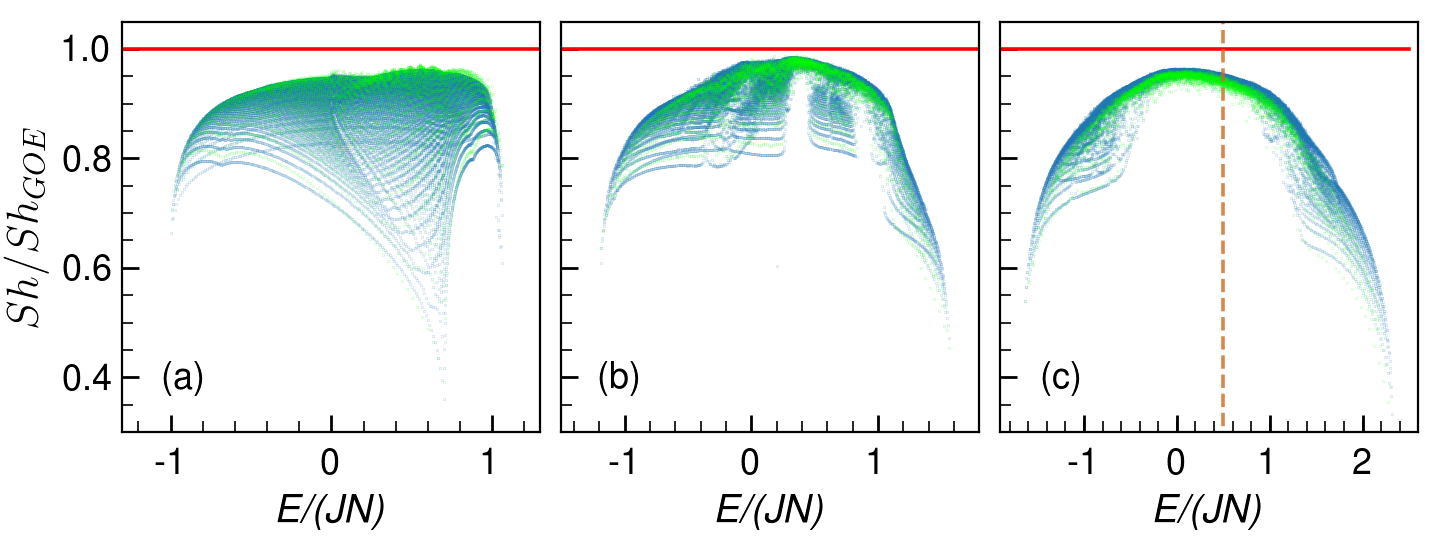} \\
	\vspace{-0.6cm}
	\includegraphics[width=1.\linewidth]{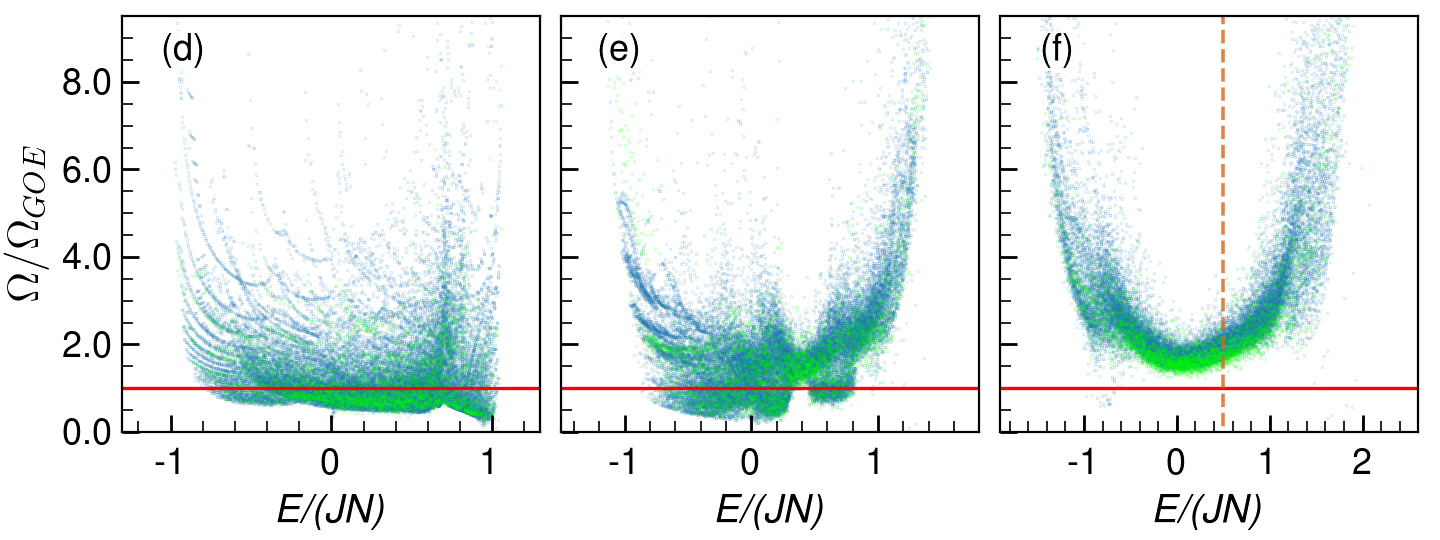}
	\caption{Shannon entropy $Sh$ and overlaps of neighboring eigenstates $\Omega$ as a function of energy for $N=90$ (light lime dots) and $N=270$ (dark blue dots). In (a) and (d):  $\epsilon/J=0$, in  (b) and (e): $\epsilon/J=0.7$, and in (c) and (f): $\epsilon/J=1.5$. The solid horizontal lines mark the results for GOE full random matrices. The dashed vertical lines in (c) and (d) mark approximately the center of the chaotic region. 
	}
		\label{fig06}
	\end{figure}

In Figs.~\ref{fig06}~(a)-(c), we show the Shannon entropy, $Sh^{\alpha}$, of each eigenstate $|\alpha \rangle $ written in the Fock basis $|n\rangle $,
\begin{equation}
Sh^{\alpha} \equiv -\sum_{n=1}^D  |C^{\alpha}_n|^2 \ln |C^{\alpha}_n|^2 ,
\label{entropyS}
\end{equation}
as a function of energy. In the equation above, $C^{\alpha}_n = \langle n | \alpha \rangle $. This entropy measures the degree of delocalization of the eigenstates in the chosen basis. If the eigenstate coincides with a basis vector, there is a single $|C^{\alpha}_n|^2 =1$ and the state is completely localized. In this case, $Sh^{\alpha}=0$. If the eigenstate is homogeneously spread in the Hilbert space, being  therefore completely delocalized, then all $|C^{\alpha}_n|^2= 1/D$ and the entropy reaches its maximum value $Sh^{\alpha}=\ln(D)$.  An equivalent measure of delocalization is the participation ratio,
\begin{equation}
PR^{\alpha} \equiv \sum_{n=1}^D  \frac{1}{|C^{\alpha}_n|^4},
\label{Eq:PR}
\end{equation}
whose figures are provided in the appendix~\ref{appPR}. The participation ratio was also considered in the analysis of the triple-well Bose-Hubbard model in Ref.~\cite{Nakerst2021}.

For GOE full random matrices, the components $C^{\alpha}_n$ of the eigenstates are independent real random variables from a Gaussian distribution with weights $|C^{\alpha}_n|^2$ that fluctuate around $1/D$, so $Sh_{\text{GOE}} \sim \ln(0.48 D)$. In Figs.~\ref{fig06}~(a)-(c), we show $Sh^{\alpha}$ divided by  $Sh_{\text{GOE}} $.

In the integrable regime [Fig.~\ref{fig06}~(a)], we see a pattern of lines that must be associated with periodic orbits likely to be found in the phase space of the classical limit of the model. This subject will be investigated in a future publication. As $\epsilon/J$ increases, regions of chaos begin to emerge [Fig.~\ref{fig06}~(b)], where the fluctuations decrease significantly and $Sh^{\alpha}$ reaches values closer to $Sh_{\text{GOE}} $, as in the vicinity of $E/(JN) \sim 0.3$ and $E/(JN) \sim 0.9$. For $\epsilon/J=1.5$ [Fig.~\ref{fig06}~(c)], an evident chaotic region emerges for $E/(JN)$ in the interval given approximately by [-0.2,1]. This energy range explains our choice for $E_c/(JN)=0.5$ in the analysis of the spectral form factor in Eq.~(\ref{Eq:rect}).

Our system is clearly separated into regions of chaos and non-chaos, independently of how large the number of bosons is. This is confirmed by comparing the results for $N=90$ (light color) and $N=270$ (dark color) in Fig.~\ref{fig06}~(c).

In Figs.~\ref{fig06}~(d)-(f), we show the quantity $\Omega_{\alpha, \alpha'}$ first proposed in Ref.~\cite{Santos2012PRE} to measure how similar two neighboring eigenstates $|\alpha \rangle$ and $|\alpha' \rangle$ are,
\begin{equation}
\Omega_{\alpha, \alpha'} \equiv \sum_{n=1}^D  |C^{\alpha}_n|^2 |C^{\alpha'}_n|^2 .
\end{equation}
In full random matrices, where the components $|C^{\alpha}_n|^2$ and $|C^{\alpha'}_n|^2$ are uncorrelated Gaussian random numbers, $\Omega_{\text{GOE}} \sim 1/D$. Correlations  result in values of $\Omega_{\alpha, \alpha'}> 1/D$. Large  values of $\Omega_{\alpha, \alpha'}$ and large fluctuations are found throughout the spectrum of the integrable model [Fig.~\ref{fig06}~(d)], while in the chaotic domain [Fig.~\ref{fig06}~(f)] they are restricted to the edges of the spectrum, $E/(JN)<-0.2$ and $E/(JN)>1$, where chaos does not develop. Notice, however, that even in the chaotic region, $\Omega >\Omega_{\text{GOE}}$, which indicates that some level of correlation among the components persists.

	\begin{figure}[h]
	\centering
	\includegraphics[width=0.9\linewidth]{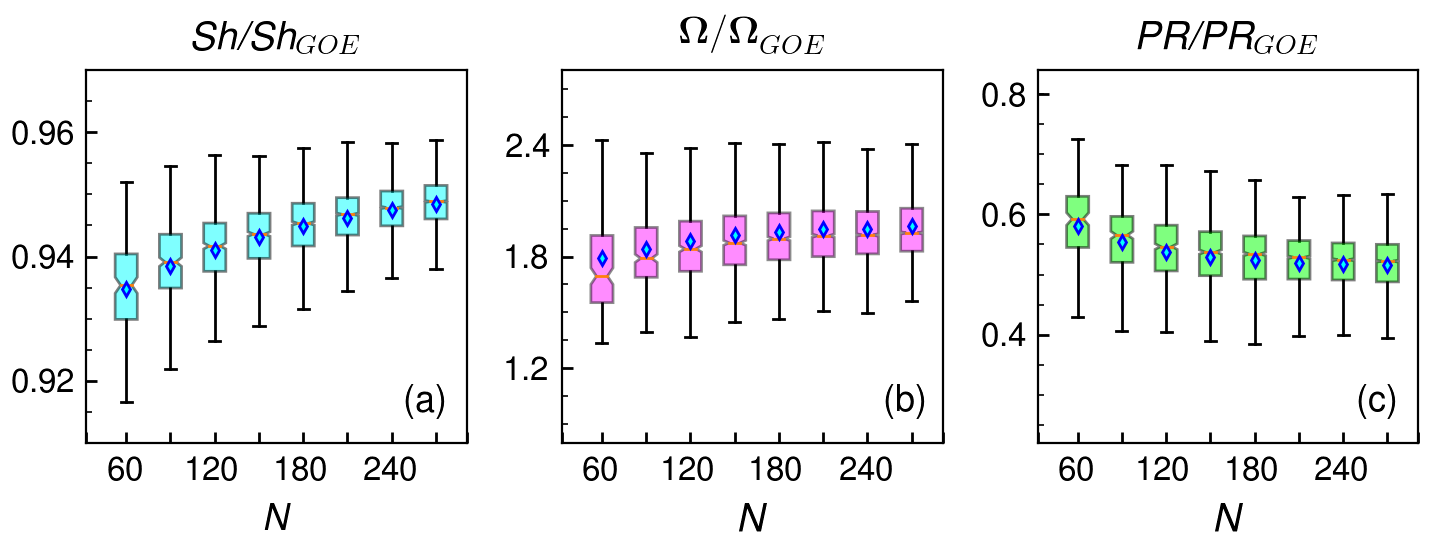}
	\caption{Box-and-whisker plots for the Shannon entropy (a), overlaps of neighboring eigenstates (b), and participation ratio (c) for various $N$'s. The data range comprises eigenstates with energies $E/(JN) \in [0.4,0.6]$.  The median for each $N$ is marked with the orange line inside each box and the average with the diamond symbol. }
		\label{boxplot}
	\end{figure}

To get some insight on how the level of correlations depend on $N$, in Figs.~\ref{boxplot}~(a)-(c), we select the eigenstates in the chaotic region with energies $E/(JN) \in [0.4,0.6]$ and study how the averages over these states for $\langle Sh \rangle/Sh_{GOE}$, $\langle \Omega \rangle/\Omega_{GOE}$, and $\langle PR\rangle/PR_{GOE}$ depend on $N$.
The analysis is done with box-and-whisker plots \cite{tukey1976exploratory}. The horizontal line drawn in the middle of the boxes indicates the median and the whiskers, which are the lines extending from the boxes, indicate the data's minimum and maximum, excluding outliers. The averages are marked with symbols.

The medians in Fig.~\ref{boxplot} suggest that as $N$ grows, the three quantities tend towards a constant. The fact that this value is below that for random matrices is understandable, since we are dealing with the eigenstates of realistic systems with two-body couplings. It calls attention, however, that the normalized averages for the entropy grows with $N$ [Fig.~\ref{boxplot}~(a)], while for $\Omega$ [Fig.~\ref{boxplot}~(b)] and $PR$ [Fig.~\ref{boxplot}~(c)], the averages move further away from the random matrix results. The overlaps of neighboring states and the participation ratio are more sensitive to fluctuations in the tails of their distributions than the Shannon entropy, due to the logarithm present in the latter~\cite{solorzano2021multifractality}.

The growth of $\langle \Omega \rangle/\Omega_{GOE}$ and the decay of $\langle PR\rangle/PR_{GOE}$ with $N$ suggest the presence of correlations and motivate investigating whether the eigenstates with energies in the most chaotic region, those with $E/(JN) \sim 0.5$, might in fact be multifractal. 

\subsection{Multifractality}

\begin{figure}[h!]
	\centering		\includegraphics[width=1\linewidth]{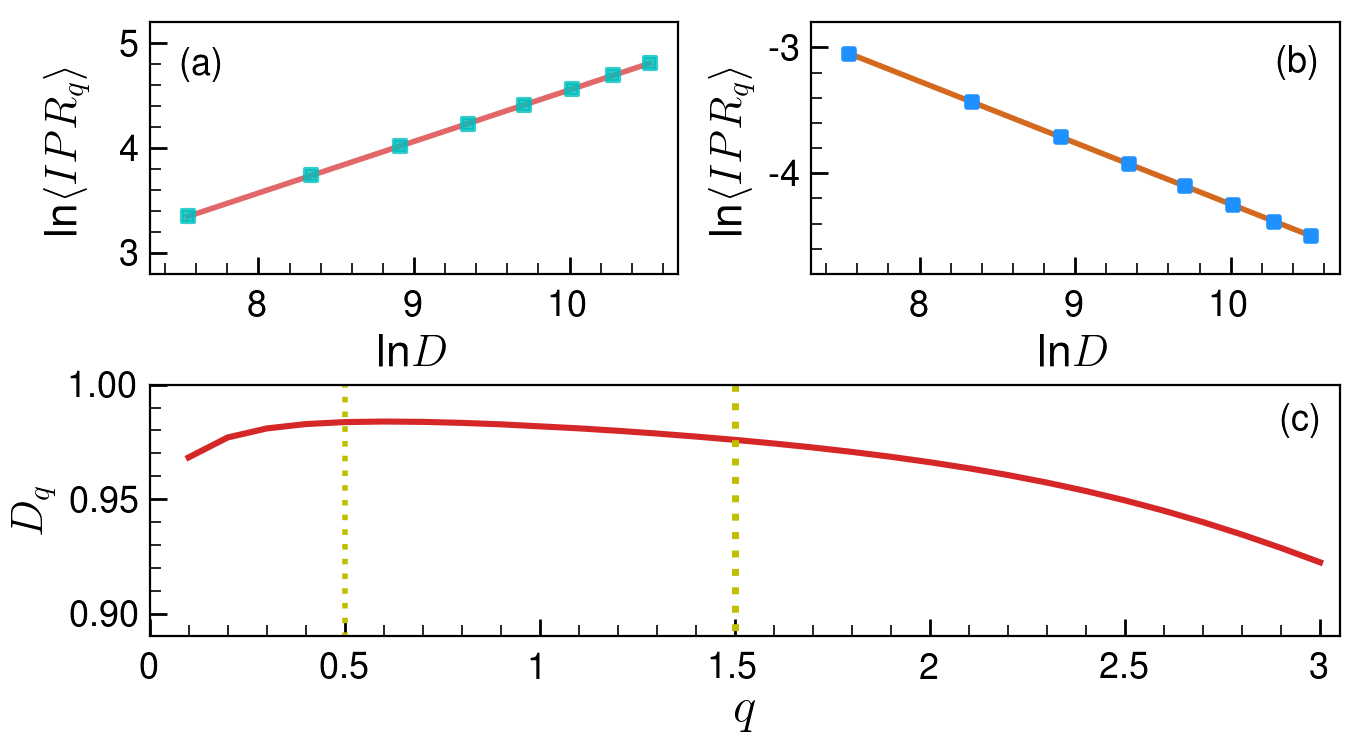} 
	\caption{Scaling analysis of the generalized inverse participation ratio averaged over 300 eigenstates with $E/(JN) \sim 0.5$ for $q=0.5$ in (a) and $q=1.5$ in (b), and generalized dimension $D_q$ as a function of $q$ in (c). In (a) and (b), the solid line is a linear fitting and the symbols are the numerical results obtained by varying the dimension of the Hilbert space from $D=1\,891$ ($N=60$) to $D=36\,856$ ($N=270$).}
		\label{figMULT}
	\end{figure}

To verify whether the eigenstates with energy $E/(JN) \sim 0.5$ are multifractal, we study how the generalized dimension $D_q$, obtained from the generalized inverse participation ratio,  $IPR_q  = \sum_n |C_n^{\alpha}|^{2q} $, depends on $q$ \cite{Wegner1980,Evers2008}. The generalized dimension is extracted from the scaling analysis of 
\begin{equation}
\langle IPR_q \rangle \propto D^{-(q-1) D_q} , 
\end{equation}
as illustrated in Fig.~\ref{figMULT}~(a) and Fig.~\ref{figMULT}~(b), where we vary the dimension of the Hilbert space from $D=1\,891$ (for $N=60$) to $D=36\,856$ (for $N=270$). A nonlinear behavior of $D_q$ as a function of $q$ indicates multifractality. This implies that the eigenstates are extended, but do not cover the entire Hilbert space ergodically, that is $0<D_q<1$.

Our results are shown in Fig.~\ref{figMULT}~(c). The values of $D_q$ are larger than 0.9, but always smaller than 1, and they are nonlinear in $q$, indicating multifractality.

\subsection{Distribution of Components}

\begin{figure}[h]
	\centering
		\includegraphics[width=1.\linewidth]{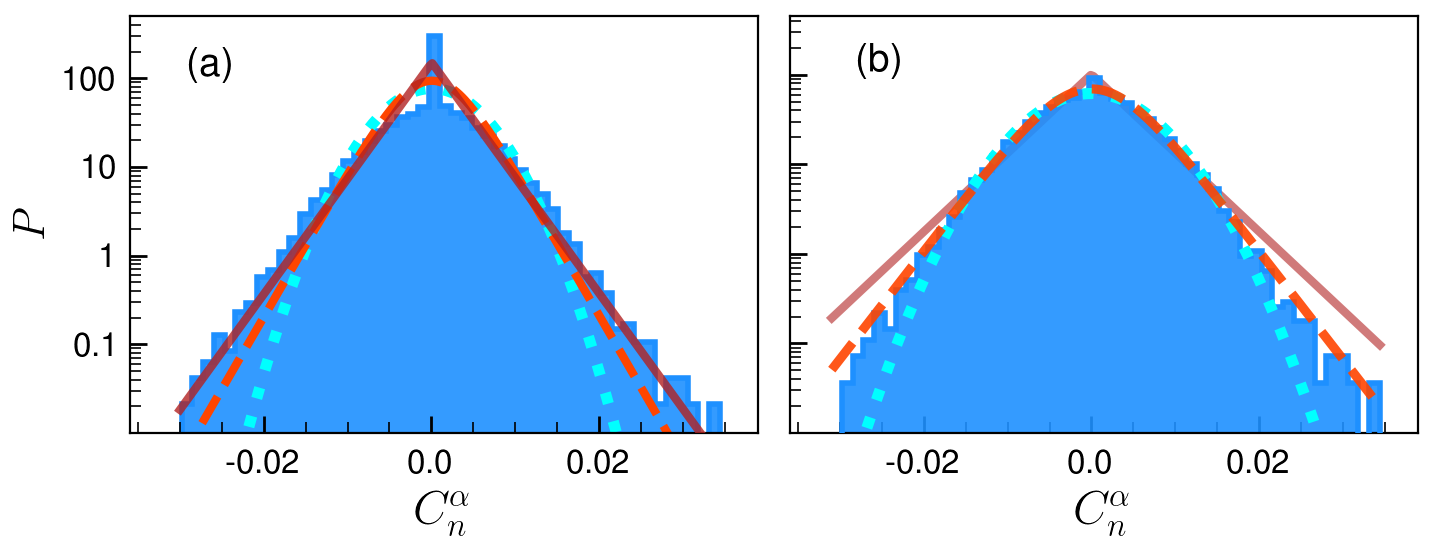}
	\caption{Distribution of the components $C_n^{\alpha}$ of an eigenstate with energy $E/(JN) \sim 0.5$; $N=270$. In (a), all components are considered, while in (b), only those for which $-0.2 \leq e_n/J \leq 1$. Solid line: Laplace distribution; dashed line:  Logistic distribution; dotted line: Gaussian distribution. }
		\label{FigCa}
	\end{figure}

The discussion above prompts a more detailed analysis of the components of the eigenstates. We select a representative eigenstate  $|\alpha \rangle = \sum_n C_n^{\alpha} |n \rangle$ with energy $E/(JN) \sim 0.5$. The distribution of its components in Fig.~\ref{FigCa}~(a) shows a high peak at $C_n^{\alpha} \sim 0$. This excessive number of zero amplitudes comes mostly from the Fock states that have energy $e_n=\langle n|H|n \rangle$ outside the chaotic region, that is $e_n/J<-0.2$ or $e_n/J>1$. By removing the components associated with these states, the peak is erased, as seen in Fig.~\ref{FigCa}~(b). The remaining Fock states constitute 59\% of the Hilbert space, but they are the main constituents of the selected eigenstate, leading to
$ \sum\limits_{\substack{n \\ -0.2 \leq e_n/J \leq 1}} \hspace{-0.4 cm} |C_n^{\alpha}|^2 = 0.90.$ 

The best distribution in Fig.~\ref{FigCa}~(a) is Laplace. After removing the peak, in Fig.~\ref{FigCa}~(b), the best distribution becomes Logistic, which is similar to Gaussian, but exhibits longer tails. This analysis makes it clear that the eigenstates of our triple-well model do not reach fully chaotic structures. 

The statement above is valid also for the triple-well Bose-Hubbard models presented in the appendix~\ref{appA}. The distributions of the components of their most delocalized eigenstates are also Logistic.

\section{Eigenstate thermalization hypothesis}

Chaotic eigenstates explain and ensure the validity of the eigenstate thermalization hypothesis (ETH)~\cite{Santos2010PRE,Santos2010PREb}. The ETH says that when the eigenstate expectation values of a few-body observable ${\cal O}$, that is $\cal{O}_{\alpha \alpha} =\langle \alpha | \hat{\cal{O}} |\alpha \rangle $, are smooth functions of the eigenenergies, these values approach the result from the microcanonical ensemble, ${\cal O}_{mic}$, as the system size increases~\cite{Dalessio2016}.  The hypothesis is also attached to the conditions of absence of degeneracies and $ \cal{O}_{\alpha \beta} \ll \cal{O}_{\alpha \alpha}$, where $\cal{O}_{\alpha \beta}=\langle \beta | \hat{O} |\alpha \rangle$ are the off-diagonal elements of the observable. These are the prerequisites for thermalization, where the infinite-time average of the observable coincides with its thermodynamic average.

 In the case of interacting many-body quantum systems, the onset of chaotic eigenstates also leads to the Gaussian distribution of the off-diagonal elements of few-body observables~\cite{Beugeling2015,foot02}. In this section, we investigate the consequences that the lack of gaussianity of the eigenstates of our model has on the diagonal and off-diagonal elements of the number operator of each well.

\subsection{Diagonal Elements} 

\begin{figure}[h]
	\centering		\includegraphics[width=1.\linewidth]{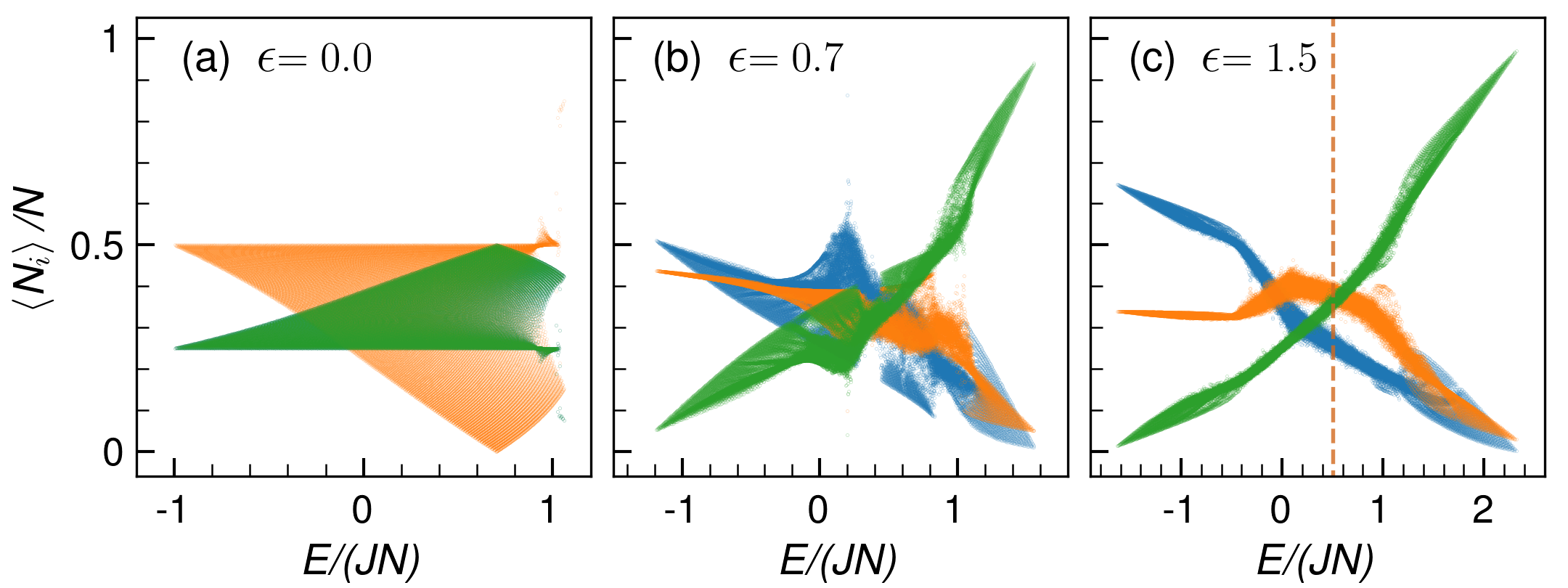}
\vspace{0.5cm}
	\includegraphics[width=0.48\linewidth]{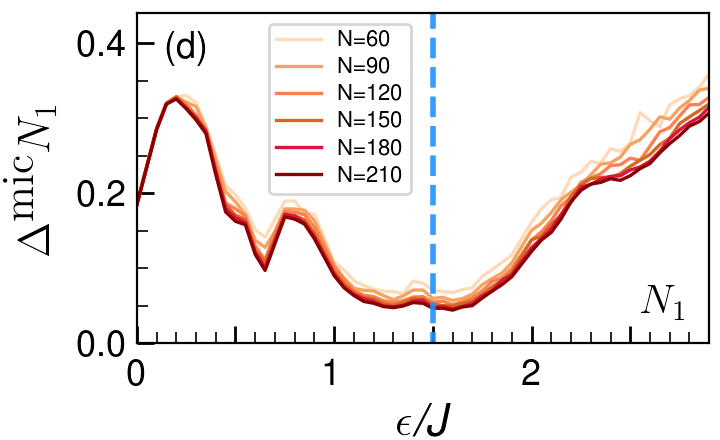}		
	\includegraphics[width=0.48\linewidth]{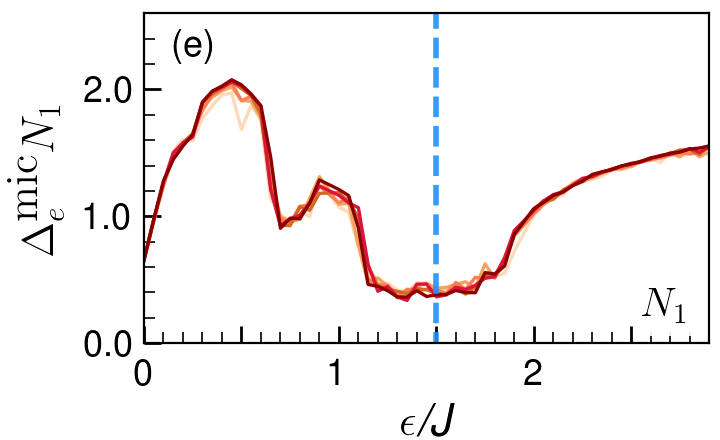}
	\caption{Eigenstate expectation values for $(N_1)_{\alpha,\alpha}$ (blue), $(N_2)_{\alpha,\alpha}$ (orange), and $(N_3)_{\alpha,\alpha}$ (green) as a function of energy, for $\epsilon/J=0.0$ (a), $\epsilon/J=0.7$ (b), and $\epsilon/J=1.5$ (c); $N=270$.
Average relative deviation of the eigenstate expectation values of $\hat{N_1}$ with respect to the microcanonical average (d) and the normalized extremal fluctuations of the eigenstate expectation values of $\hat{N_1}$ (e) both as a function of the integrability breaking term $\epsilon$. In (d) and (e), the eigenstates lie in the  energy range 
$[E/(JN)-\Delta E/(JN),E/(JN)+\Delta E/(JN)]$ with $E/(JN)=0.5$ and $\Delta E/(JN)=0.1$.}
		\label{fig07}
	\end{figure}

We start the analysis by investigating the diagonal elements of $\hat{N}_i$ in Fig.~\ref{fig07}. As the integrability breaking term increases from $\epsilon/J =0$  in Fig.~\ref{fig07}~(a) to $\epsilon/J =1.5$ in  Fig.~\ref{fig07}~(c), the fluctuations decrease significantly, reflecting the similar behavior of the eigenstates illustrated in Fig.~\ref{fig06}. For the integrable model in Fig.~\ref{fig07}~(a), there is a clear regular structure, and $\langle N_1 \rangle = \langle N_3 \rangle$ due to the Hamiltonian parity symmetry. In Fig.~\ref{fig07}~(b), smaller fluctuations appear for $E/(JN) \sim 0.3$ and $E/(JN) \sim 0.9$, as it happens also for the entropy in Fig.~\ref{fig06}~(b). In Fig.~\ref{fig07}~(c), smaller fluctuations are seen throughout the spectrum, although outside the chaotic region, for $E/(JN)<-0.2$ and $E/(JN)>1$, the pattern of lines seen for the eigenstates in Fig.~\ref{fig07}~(c) are carried over for the  expectation values of the number operators.

Close to $E/(JN) \sim 0.5$ in Fig.~\ref{fig07}~(c), the population inversion, where $(N_2)_{\alpha \alpha}$ (orange) and  $(N_3)_{\alpha \alpha}$ (green)  become larger than  $(N_1)_{\alpha \alpha}$ (blue), is consistent with the tilt, which causes states with occupation on site 2 and, especially, on site 3 to have larger energies than states with population on site 1. For very high energies, it is therefore natural that $\langle N_1\rangle \simeq \langle N_2\rangle \to 0$. In contrast, for low energies, the distribution of particles is relatively symmetric around  $\langle N_2\rangle$, with $N_1>N_2>N_3$ and $N_3 \to 0$, as expected.

To study the fluctuations of an observable around the microcanonical expectation value, we consider the deviation of its eigenstate expectation value, 
\begin{equation}
    \Delta^{mic}\mathcal{O} =  \frac{\sum_\alpha |\mathcal{O}_{\alpha\alpha}-\mathcal{O}_{mic}|}{\sum_\alpha\mathcal{O}_{\alpha\alpha}},
    \label{micro}
\end{equation}
with respect to the microcanonical result,
 \begin{equation}
    \mathcal{O}_{mic} =  \frac{1}{\mathcal{N}}_{E,\Delta E} \sum_{\substack{\alpha,\\|E-E_\alpha|<\Delta E}} \mathcal{O}_{\alpha\alpha},
\end{equation}
where ${\mathcal{N}}_{E,\Delta E}$ is the number of energy eigenstates with energy in the window $\Delta E$.
We also study the  normalized extremal fluctuation ~\cite{Santos2010PREb},
\begin{equation}
     \Delta_e^{mic}\mathcal{O} = |\frac{\text{max}\, \mathcal{O}-\text{min} \, \mathcal{O}}{\mathcal{O}_{mic}}|,
         \label{microe}
 \end{equation}
 where $\text{max} \, \mathcal{O}$ and $\text{min} \,\mathcal{O}$ are the maximum and minimum values of $\mathcal{O}_{\alpha\alpha}$.
In Fig.~\ref{fig07}~(d) [Fig.~\ref{fig07}~(e)], we present the results for $\Delta^{mic} N_1$ [$\Delta^{mic}_e N_1$] for eigenstates with $E/(JN)=0.5$ in the window of width $\Delta E/(JN)=0.1$. The results for $(N_2)_{\alpha \alpha}$ and $(N_3)_{\alpha \alpha}$ are similar (not shown).

Figures~\ref{fig07}~(d)-(e) are analogous to Fig.~\ref{fig04}. They show that the smallest fluctuations of the eigenstate expectation values happen for $\epsilon/J \sim 1.5$, where the chaos indicator $\beta$ is also the largest. The fluctuations increase as the system approaches both integrable limits, as $\epsilon/J \rightarrow 0$ (Bethe ansatz) and as $\epsilon/J \rightarrow \infty$ (self-trapping).
	
At a fixed value of $\epsilon/J$, one sees that $\Delta^{mic} N_1$ in Fig.~\ref{fig07}~(d) decreases slightly as the total number  of particles increases. A discussion of how $\Delta^{mic} {\cal O}$ scales with the dimension $D$ of the Hilbert for the triple-well Bose-Hubbard model is provided in Ref.~\cite{Nakerst2021}, where it is found that the scaling does not follow expectations consistent with fully chaotic eigenstates. Similarly to our analysis of Fig.~\ref{fig04}, the results in Fig.~\ref{fig07}~(d) suggest that the reduction of the fluctuations for larger $N$'s is caused by better statistics, not necessarily improved levels of chaos.

Our results for the extremal fluctuations in Fig.~\ref{fig07}~(e) add to the above discussion. We see that $\Delta^{mic}_e N_1$ does not decrease as $N$ increases, at least not for the numbers that we consider. This contrasts with the case of interacting many-body quantum systems with many sites, where the extremal fluctuations do decrease as the system size increases. The extremal fluctuation is a more rigorous test of the validity of the ETH~\cite{Santos2010PREb}, and by extension of the degree of quantum chaos. Figure~\ref{fig07}~(e) is a compelling indication that the level of chaoticity of the model does not improve by increasing $N$.

\subsection{Off-diagonal elements}

The strongest signatures of quantum chaos for our triple-well model happen for $\epsilon/J \sim 1.5$, but the results for level statistics [Fig.~\ref{fig04}], structure of the eigenstates [Fig.~\ref{FigCa}], and extremal fluctuations [Fig.~\ref{fig07}~(e)] indicate that even at this point, full chaos is not achieved. Here, we investigate how this saturated level of chaos, in particular the non-Gaussian distribution of the eigenstates components in Fig.~\ref{FigCa}, gets reflected into the distribution of the off-diagonal elements of the number  operators.

The off-diagonal elements of $\hat{N}_i$ is given by
\begin{eqnarray}
&&\langle \alpha | \hat{N}_i |\beta \rangle = \sum_{n=1}^D  C_n^\alpha C_n^\beta \langle n | \hat{N}_i |n \rangle \nonumber \\
&& 
= \hspace{-0.4 cm}
\sum_{\substack{n=1\\\langle n| \hat{N}_i |n \rangle = 1 }}^N \hspace{-0.4 cm} C_n^\alpha C_n^\beta + 2 \hspace{-0.4 cm} \sum_{\substack{n=1\\\langle n| \hat{N}_i |n \rangle = 2 }}^{N-1}  \hspace{-0.4 cm} C_n^\alpha C_n^\beta + \ldots +N \hspace{-0.4 cm} \sum_{\substack{n=1\\\langle n| \hat{N}_i |n \rangle = N }}^{1} \hspace{-0.4 cm} C_n^\alpha C_n^\beta .\nonumber \\ 
&&\label{Eq:off}
\end{eqnarray}
In the case of fully chaotic eigenstates, where $C_n^{\alpha}$'s are independent Gaussian random numbers, the distribution of $(N_i)_{\alpha \beta}$ should also be Gaussian. This is evident from the equation above. The product of independent random variables is again an independent random variable, and according to the central limit theorem, the sum of random variables from any distribution follows a Gaussian distribution.

\begin{figure}[h]
	\centering
		\includegraphics[width=1.\linewidth]{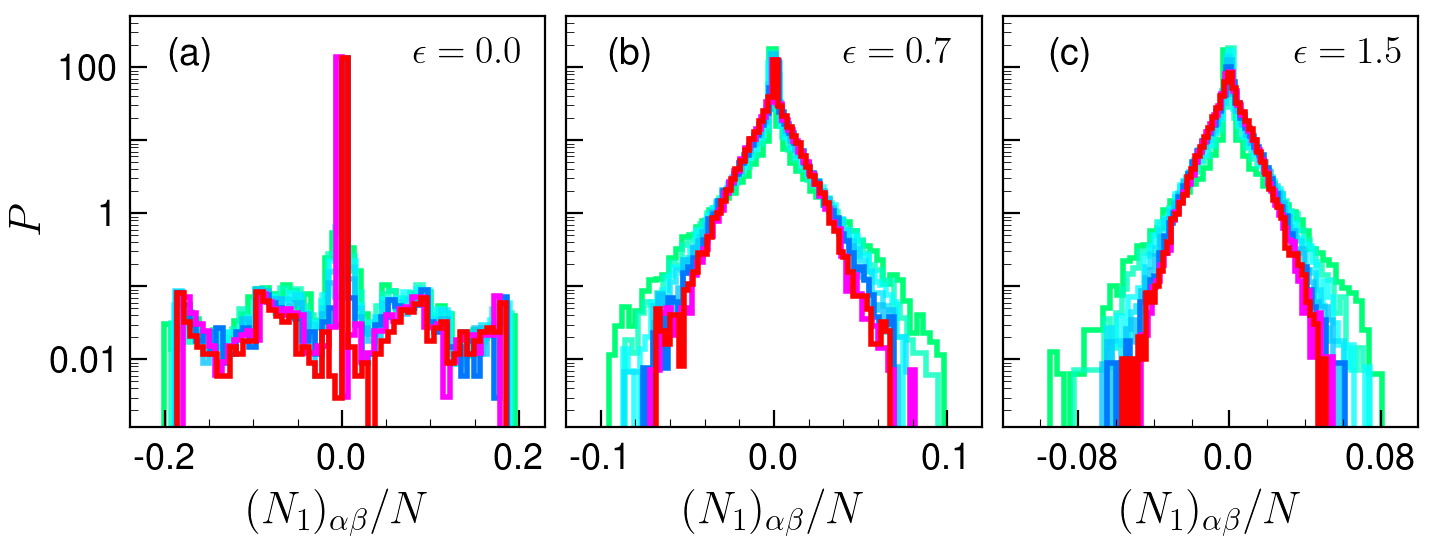}	
	\caption{Distributions of the off-diagonal elements of the number operator of well 1, $\langle \alpha | \hat{N}_1 |\beta \rangle $, for 300 eigenstates with energy $E/(JN) \sim 0.5$. The value of the integrability breaking parameter is indicated in the panels. The distributions are shown for different numbers of particles, $N=60$ (green) to $N=270$ (red) in increments of 30. }
		\label{fig11}
	\end{figure}

In Fig.~\ref{fig11}, we show the distribution of the number operator of well 1 (for equivalent results for wells 2 and 3, see the appendix~\ref{appETH}.)
As the integrability term $\epsilon/J$ increases from zero [Fig.~\ref{fig11}~(a)] to 1.5 [Figs.~\ref{fig11}~(c)], the peak at $(N_1)_{\alpha \beta}/N \sim 0$ decreases and the distribution gets more similar to a Gaussian, although this shape is never achieved, independently of the number of particles. 

\begin{figure}[h]
	\centering
	\includegraphics[width=1.\linewidth]{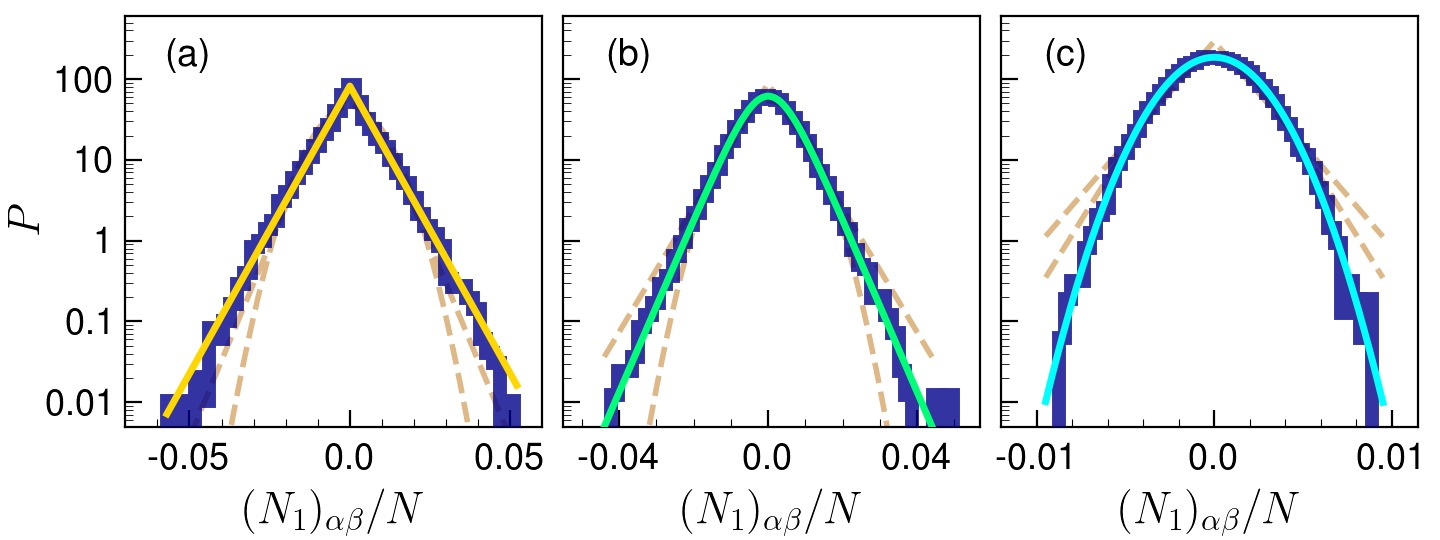} 
	\caption{Distributions of the off-diagonal elements of the number operator of well 1, $\langle \alpha | \hat{N}_1 |\beta \rangle, $ for 300 eigenstates with $E/(JN)\sim 0.5$, $N=270$,  $\epsilon/J=1.5$. The solid line indicates the best fit: (a) Laplace, (b) Logistic, and (c) Gaussian distribution.
	In (a), just as in Fig.~\ref{fig11}~(c), all the components $C_n^{\alpha}$ of the eigenstates are considered. In (b), only those components for which $0.25 \leq e_n/J \leq 0.7$ are taken into account. In (c), the components are those from Gaussian random vectors. 
	}
	\label{Fig:OFF270}
	\end{figure}

In Fig.~\ref{Fig:OFF270}~(a), we select only the curve for $N=270$ from Fig.~\ref{fig11}~(c) and show that its best fit is a Laplace distribution. Some explanations are now in order. The Laplace distribution (more precisely, a modified Bessel function of the second kind) describes the off-diagonal elements of single-particle eigenstates in chaotic quadratic Hamiltonians~\cite{RigolVidmar}. In this case, $N=1$ and the only term that survives in Eq.~(\ref{Eq:off}) is the last one. This term is a single  product of two Gaussian random variables, whose distribution is indeed Laplace. Our scenario is completely different from this one, since in Eq.~(\ref{Eq:off}), we have large sums of the products $C_n^\alpha C_n^\beta$.

Similar to our analysis in Fig.~\ref{FigCa}, a closer study of  Fig.~\ref{Fig:OFF270}~(a) reveals that the peak at $(N_1)_{\alpha \beta}/N \sim 0$ is caused by the Fock states with  energies outside the chaotic region. By removing the contributions from the states with $e_n/J<0.25$ and $e_n/J > 0.75$, the distribution of $(N_1)_{\alpha \beta}$ becomes Logistic, as seen in Fig.~\ref{Fig:OFF270}~(b), which is  closer but not yet Gaussian. If, however, we calculate Eq.~(\ref{Eq:off}) using eigenstates from GOE random matrices, then we finally reach the Gaussian shape, as expected from the central limit theorem and as illustrated in Fig.~\ref{Fig:OFF270}~(c).

The study of the off-diagonal elements corroborates our claims that the triple-well model in Eq.~(\ref{QH}) do not have fully chaotic eigenstates. The same holds for the triple-well Bose-Hubbard models presented in the appendix~\ref{appA}, where the distributions of the off-diagonal elements of the number operators are not Gaussian either.

\section{Conclusions}
We investigated the spectrum, eigenstates, and occupation numbers of an integrable bosonic triple-well model that becomes chaotic with the addition of a tilting potential. The analysis of the structure of the eigenstates shows that for values of the tilt where chaos emerges, there are still regions of energy where the system remains non-chaotic. Furthermore, even within the energy interval of chaos, the eigenstates are not fully chaotic, that is, their components do not follow Gaussian distributions and the generalized dimensions are smaller than 1, which suggest reminiscences of correlations. The distributions of the off-diagonal elements of the number operators are particularly sensitive to the lack of gaussianity of the eigenstates, which prevents those distributions from becoming Gaussian. 

Studies of the eigenstates and off-diagonal elements of observables can reveal details about quantum systems that are not always easily accessible from a direct study of their eigenvalues. In our specific case, the analysis of the eigenstates and observables shows that three wells constitute the preface for many-body quantum chaos. A natural extension of our work is therefore to examine how our results change by increasing the number of wells. It is also worth investigating the role played by the geometry of the system and by the addition of nonlinear terms~\cite{Rubeni2017} or external drives~\cite{Kidd2019, Kidd2020}.

\begin{acknowledgments}
K.W.W, E.R.C., and A.F. are grateful to the Brazilian agency CNPq (Conselho Nacional de Desenvolvimento Cient\'ifico e Tecnol\'ogico) for partial financial support. K.W.W. also acknowledges support from
Sociedade Brasileira de Física (SBF)/American Physical Society (APS) through a Brazil-US Physics Student Visit Program.
L.F.S. is supported by the United States National Science Foundation (NSF, Grant No. DMR-1936006). We thank Thomas Wittmann Wilsmann for helpful discussions.
\end{acknowledgments}


\appendix

\section{Bose-Hubbard Models}
\label{appA}

Bose-Hubbard models describe interacting spinless bosons on a discrete lattice~\cite{Fisher1989} and are experimentally implemented with ultracold atoms in optical lattices~\cite{Bloch2005}. In the case of three wells, the bare Bose-Hubbard model is represented by the Hamiltonian
\begin{align}
	\label{BHbare}
	\hat{H} = &\frac{U_0}{N}\left(\hat{N}_1\left(\hat{N}_1-1\right)+\hat{N}_2\left(\hat{N}_2-1\right)+\hat{N}_3\left(\hat{N}_3-1\right)\right) \nonumber \\
 &+\frac{J}{\sqrt{2}}\left(\hat{a}_1^\dagger \hat{a}_2 + \hat{a}_2^\dagger \hat{a}_1\right)+\frac{J}{\sqrt{2}}\left(\hat{a}_2^\dagger \hat{a}_3 + \hat{a}_3^\dagger \hat{a}_2\right),
\end{align}
where $U_0$ is the onsite interaction, $J$ is the hopping (tunneling) parameter, and $N=N_1+N_2+N_3$ is the total number of particles.

This system presents signatures of quantum chaos when the number $L$ of wells coincides with the number particles, $L=N \geq 5$ \cite{Kolovsky2004}. However, as shown in~\cite{Nakerst2021}, the model is also chaotic for only 3 sites and $N \gg3$. Notice that the Hamiltonian has parity symmetry, so to study level statistics, one should either break this symmetry, as done in~\cite{Nakerst2021}, or separate the eigenvalues by symmetry sector. An alternative is to resort to the correlation hole, which detects level repulsion even in the presence of symmetries~\cite{Cruz2020,Santos2020}.

The extended version of the Bose-Hubbard model,
\begin{align}
\label{BH}
	\hat{H} = &\frac{U_0}{N}\left(\hat{N}_1\left(\hat{N}_1-1\right)+\hat{N}_2\left(\hat{N}_2-1\right)+\hat{N}_3\left(\hat{N}_3-1\right)\right) \nonumber \\
	&+ \frac{U_1}{N}\left(\hat{N}_1\hat{N}_2+\hat{N}_2\hat{N}_3+ \frac{1}{\alpha}(\hat{N}_1\hat{N}_3)\right) \nonumber \\ &+\frac{J}{\sqrt{2}}\left(\hat{a}_1^\dagger \hat{a}_2 + \hat{a}_2^\dagger \hat{a}_1\right)+\frac{J}{\sqrt{2}}\left(\hat{a}_2^\dagger \hat{a}_3 + \hat{a}_3^\dagger \hat{a}_2\right),
\end{align}
includes also interactions between the wells, which emerge in dipolar gases. As discussed in~\cite{Lahaye2010}, the parameter $\alpha$ depends on the geometry of the trap and
can vary between $4\leq \alpha \leq 8$. The extended Bose-Hubbard model also has parity symmetry through exchange of wells 1 and 3. Depending on the choices of parameters and with some rearrangement of the signs, Eq.~(\ref{BH}) coincides with the Hamiltonian of our model in Eq.~(\ref{QH}) in the integrable limit. 

\begin{figure}[h]
	\centering
	\includegraphics[width=1.\linewidth]{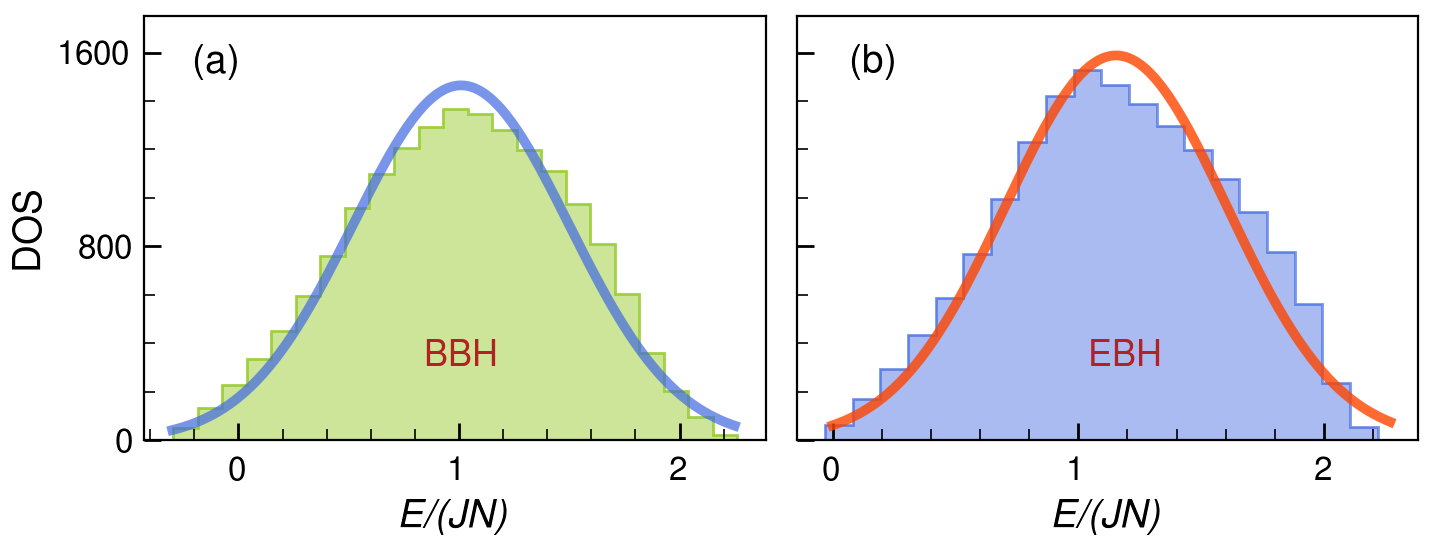}
	\caption{Density of states for (a) the bare Bose-Hubbard model from Eq.~(\ref{BHbare})  and (b) the extended Bose-Hubbard model from Eq.~(\ref{BH}). All eigenvalues from both symmetry sectors are considered;  $N=180$. The parameters used  lead to Wigner-Dyson distributions: (a) $U_0/J=2.03$, (b) $U_0/J=1.85$, $U_1/J=1.2$ and $\alpha=2\sqrt{2}$. The solid lines are Gaussian fits.
	}
	\label{DOS-BHbare}
\end{figure}

In Fig.~\ref{DOS-BHbare}, we show the DOS for the two Bose-Hubbard models above for parameters that lead to approximate Wigner-Dyson distributions. For $N=180$ and the positive parity sector, we get Brody factors  $\beta\approx 0.8$. Figure~\ref{DOS-BHbare} can be compared with the DOS for the chaotic triple-well model with the external tilt in Fig.~\ref{fig03}~(c). None of the distributions, that in Fig.~\ref{fig03}~(c) or the ones in Fig.~\ref{DOS-BHbare}, have a Gaussian shape. 

\begin{figure}[h]
	\centering
	\includegraphics[width=1.\linewidth]{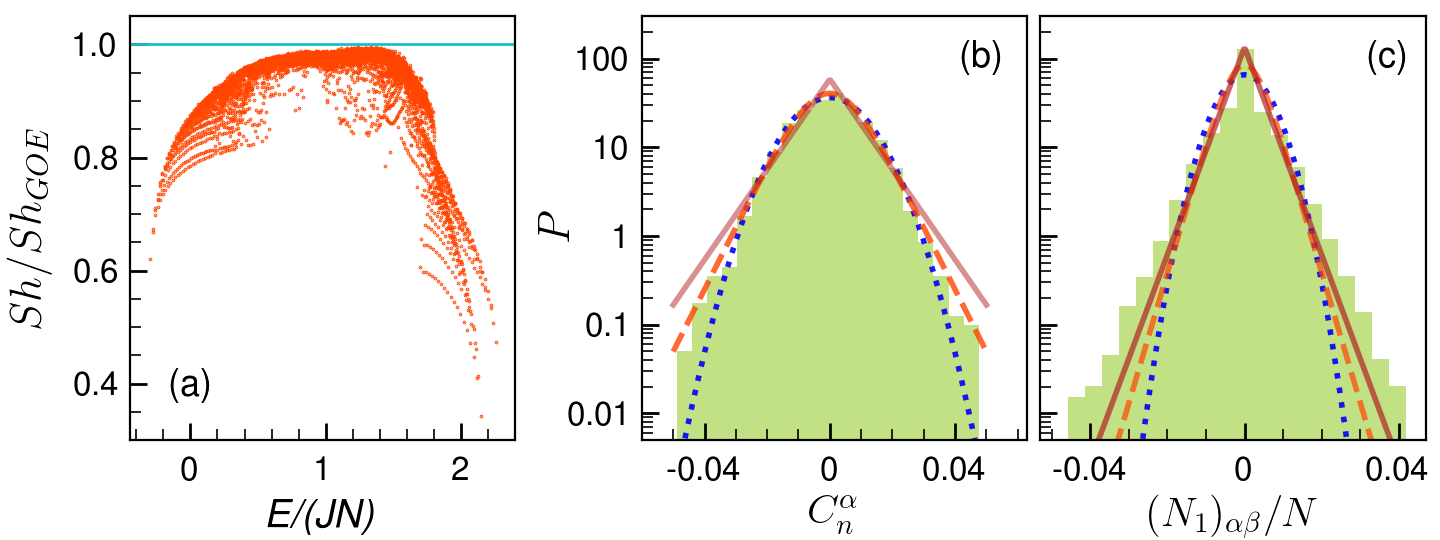}\\
		\vspace{-0.6cm}
		\includegraphics[width=1.\linewidth]{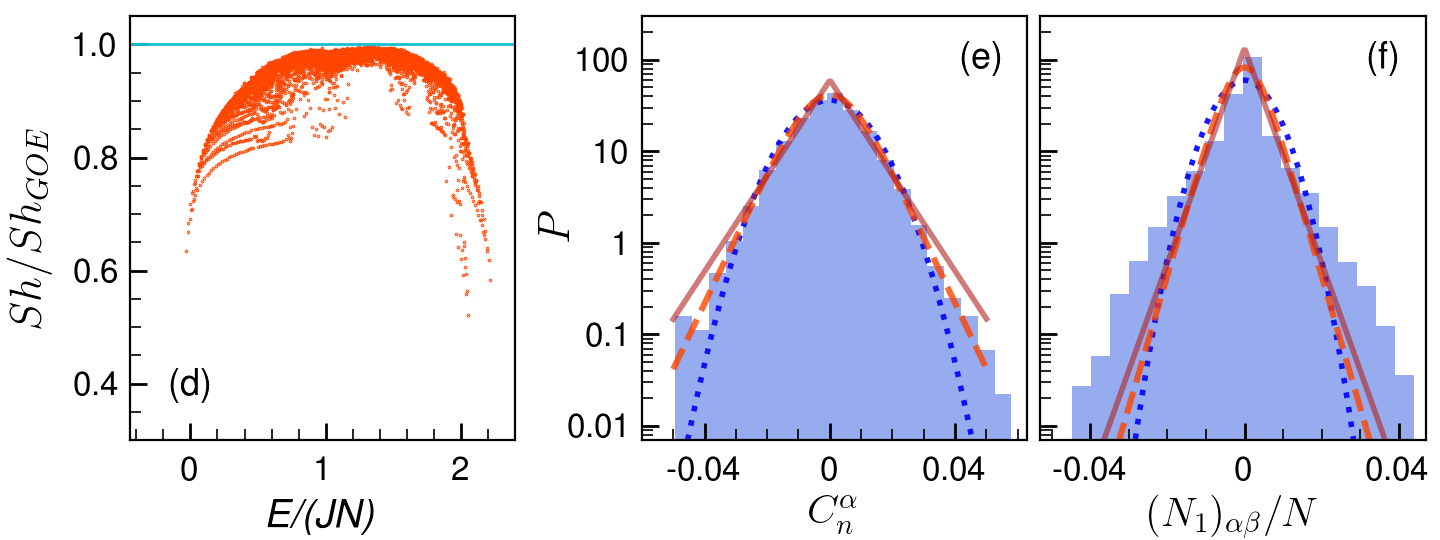}
	\caption{Shannon entropy (a,d), distribution of the components of an eigenstate in the chaotic region (b,e), and distribution of the off-diagonal elements of the number operator of well 1 (c,f) for the bare Bose-Hubbard model (a)-(c) and the extended Bose-Hubbard model (d)-(f) in the positive parity sector; $N=180$. The parameters are the same as in Fig.~\ref{DOS-BHbare}. In (a,d): Solid line indicates the result for random matrix theory. In (b,c,e,f): Solid line represents the Laplace; dashed line, the Logistic; and dotted line, the Gaussian distribution. The best fit in (b) and (e) is the Logistic distribution. In (b), $E/(JN)=1.16$ and in (e), $E/(JN)=1.33$. 
	}
	\label{OFF-BH}
\end{figure}

In Fig.~\ref{OFF-BH}, we show results for the Shannon entropy, components of the eigenstates, and off-diagonal elements of $\hat{N}_1$ for both Bose-Hubbard models in the chaotic domain. The plot for the Shannon entropy in Figs.~\ref{OFF-BH}~(a,d) can be compared with Fig.~\ref{fig06}. Similarly to our model, the Bose-Hubbard models present a region of energy away from the edges of the spectrum where the entropy is larger and has smaller fluctuations. As we move closer to borders of the spectrum, a pattern of regular lines similar to those in Fig.~\ref{fig06} appear.

We studied the distributions of the components of various eigenstates in the chaotic region of the spectrum, with energy $E/(JN)\sim1$ [$E/(JN)\sim1.4$] for the bare Bose-Hubbard model [extended Bose-Hubbard model]. In most cases, the best fit is a Logistic distribution, as illustrated in Fig.~\ref{OFF-BH}~(b) [Fig.~\ref{OFF-BH}~(e)]. For the Bose-Hubbard models, we do not find an excessive number of $C_n^{\alpha} \sim 0 $ as in Fig.~\ref{FigCa}~(a), but the tails are still longer than in Gaussian distributions.

The lack of gaussianity of the eigenstates result in the non-Gaussian distributions of the off-diagonal elements of the number operators. This is illustrated in Fig.~\ref{OFF-BH}~(c) and Fig.~\ref{OFF-BH}~(f) for $\hat{N}_1$. Contrary to Fig.~\ref{Fig:OFF270}, none of the usual distributions, Laplace, Logistic, Gaussian, or Lorentzian, capture well the histogram for $(N_1)_{\alpha \beta}$. 

\section{Additional results for our triple-well model}

We leave to this appendix some further details about our triple-well model. This includes the dependence of the timescales of the spectral form factor on $N$, a plot for the participation ratio, and the distributions of off-diagonal elements for the number operators of the three wells.

\subsection{Timescales for the spectral form factor}
\label{appFORM}

The time $t_\text{min}$ to reach the minimum of the correlation hole and the time for the saturation of the spectral form factor (Heisenberg time) can be derived using the analytical expression in Eq.~(\ref{Eq:analytics}).

\subsubsection{Time for the minimum of the correlation hole}

To determine the time $t_\text{min}$, we consider the envelope of the initial oscillatory decay, $\sin^2(\sigma t) \to 1$, (the choice $\sin^2(\sigma t) \to 1/2$ would also be suitable) and use the expression of the function $b_2(t/(2 \pi \nu_c))$ for short times, $t\leq 2 \pi \nu_c$. The latter is justified, because the minimum of $S_{FF}(t)$ is the point where the function $ [\eta \sin^2(\sigma t)]/(\sigma t)^2 $, that causes the decay of the spectral form factor, meets the $b_2(t/(2 \pi \nu_c))$ function, which is responsible for bringing $S_{FF}(t)$ up to saturation.  The time  is then obtained from 
	\begin{eqnarray}
		& &\frac{dS_{FF}^{analyt}}{dt}=0 \\
		&& \frac{4\eta}{\sigma^2 t_{\mathrm{min}}^3}=\frac{1}{\pi\nu_c}+\frac{1}{\pi\nu_c(1+\frac{t_{\mathrm{min}}}{\pi\nu_c})}-\frac{\ln\left(1+\frac{t_{\mathrm{min}}}{\pi\nu_c}\right)}{\pi\nu_c}, \nonumber
	\end{eqnarray}
which can be solved numerically to determine $t_\text{min}$.
By expanding the equation above, using $t_\text{min} \ll \pi \nu_c$, we get that
\begin{equation}
t_{\text{min}} = \left(\frac{2\pi \nu_c \eta}{\sigma^2}   \right)^{1/3} =  
\left(\frac{16 \pi  }{9 \langle \overline{S_{FF}}  \rangle^2 \sigma^3 }   \right)^{1/3} .
\end{equation}
Since $\langle \overline{S_{FF}}  \rangle$ scales with the inverse of the dimension of the Hilbert space, that is  $\langle \overline{S_{FF}}  \rangle \propto N^{-2}$, and $\sigma \propto N$, we have that  $t_{\text{min}}$ grows with the number of particles as
\begin{equation}
    t_{\text{min}} \propto N^{1/3}.
\end{equation} 
This is confirmed numerically for all $N$'s considered here, as indicated by the values of $t_{\text{min}}$ marked with circles in  Fig.~\ref{fig05}.

\subsubsection{Saturation time}

The saturation time, $t_{\mathrm{S}}$, corresponds to the time when $S_{FF}(t)$ reaches its infinite-time average  $\langle \overline{S_{FF}}  \rangle$. At these very long times, only the $b_2$ function is relevant, and since it shows a power-law behavior, $ b_2 \left( \dfrac{t}{2 \pi \nu_c} \right) \rightarrow \dfrac{\pi^2 \nu_c^2}{3 t^2}$, the complete saturation is not well determined~\cite{Schiulaz2019}. We define $t_{\mathrm{S}}$ as the moment when $S_{FF}(t_{\mathrm{S}})=(1-\delta)\langle \overline{S_{FF}}  \rangle$, where $\delta$ is a small value that guarantees  that $S_{FF}(t)$ is already within the fluctuations around the infinite-time average. This gives
\begin{equation}
		\frac{t_{\mathrm{S}}}{2\pi \nu_c}\ln\left(\frac{t_{\mathrm{S}}/\pi\nu_c +1}{t_{\mathrm{S}}/\pi\nu_c - 1}\right)=\delta\frac{(\eta-1)\langle \overline{S_{FF}} \rangle}{1-\langle \overline{S_{FF}} \rangle}+1
	\end{equation}
and using that $t_\text{S} \gg \pi \nu_c$, we arrive at
\begin{equation}
 t_{\mathrm{S}}=\frac{\pi \nu_c}{2 \sqrt{\delta}}  \propto N, 
\end{equation}
which shows that the saturation time grows linearly with $N$, as confirmed in  Fig.~\ref{fig05}, where $t_{\mathrm{S}}$ is marked with diamonds.

It is instructive to compare   $t_{\text{min}}$ and $t_{\text{S}}$  for our model with the same timescales for the Dicke model~\cite{Lerma2019}, which has two degrees of freedom, and for the one-dimensional disordered spin-1/2 model with many excitations~\cite{Schiulaz2019}, which has many degrees of freedom and a Hilbert space that grows exponentially with the number of sites. While for our  model and the Dicke model, $t_{\text{min}}$ scales with the number of particles as $N^{1/3}$ and $N^{1/2}$, respectively, and $t_{\text{S}} \propto N$, for the interacting many-body spin system, $t_{\text{min}}$ grows with the size of the Hilbert space as $D^{2/3}$ and  $t_{\text{R}} \propto D$. Based on these timescales, it might be possible to detect the correlation hole experimentally with the triple-well model, but more unlikely to get this done with many-body systems with many sites and short-range couplings.

\subsection{Participation Ratio}
\label{appPR}

\begin{figure}[h]
	\centering	
	\vskip 0.2 cm \includegraphics[width=1\linewidth]{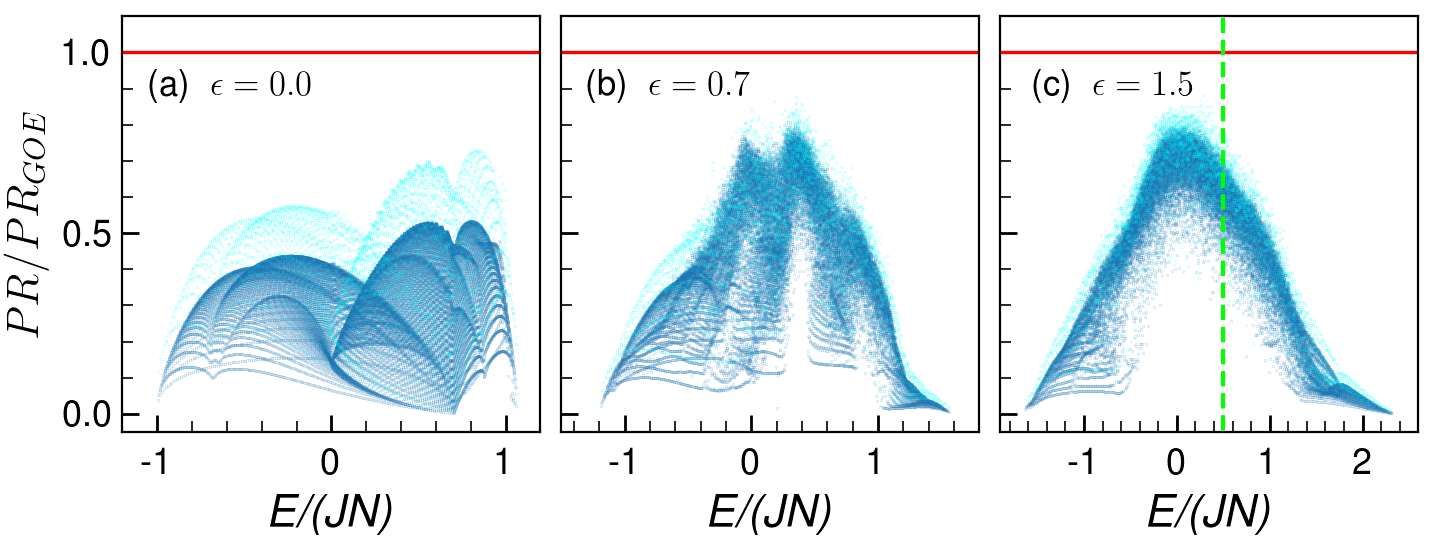} 
	\caption{Participation ratio as a function of energy for $N=90$ (light cyan dots) and $N=270$ (dark blue dots).  The solid horizontal lines mark the results for GOE random matrices. The dashed vertical line in (c)  marks approximately the center of the chaotic region.}
		\label{figPR}
	\end{figure}

We show in Fig.~\ref{figPR} the participation ratio divided by the result for GOE random matrices, $PR_{\text{GOE}} \sim  D/3$. In comparison to the results for the Shannon entropy presented in Fig.~\ref{fig06}, we  see that the fluctuations are larger for the participation ratio. 

The fluctuations decrease as the system moves from the integrable limit of Fig.~\ref{figPR}~(a) to the chaotic domain of Fig.~\ref{figPR}~(c), but even for $\epsilon=1.5$, we still find regions closer to the edges of the spectrum with patterns of lines similar to those found in the regular regime. In addition, the participation ratio is throughout smaller than $PR_{\text{GOE}}$ and this does not improve as $N$ increases [cf. $N=270$ (dark dots) with $N=90$ (light dots)].

\subsection{Distributions of off-diagonal elements}
\label{appETH}

\begin{figure}[h]
	\centering
	\includegraphics[width=1.\linewidth]{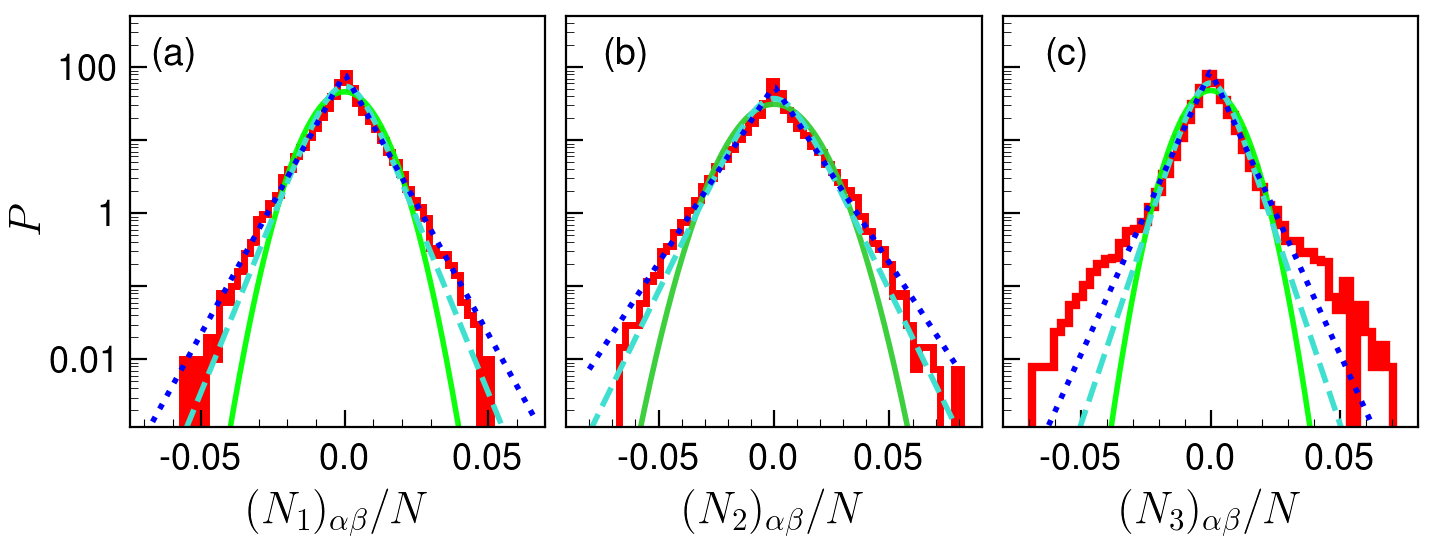} 
	\caption{Distributions of the off-diagonal elements of (a) $\hat{N}_1$, (b) $\hat{N}_2$, and (c) $\hat{N}_3$  for 300 eigenstates with energy $E/(JN) \sim0.5$; $N=270$, $U/J=0.7$ and $\epsilon/J=1.5$.  The fitting curves correspond to  Laplace (dashed line), Logistic (dashed line) and Gaussian (solid line) distributions.
	}
	\label{Fig:OFFmore}
	\end{figure}

In Fig.~\ref{Fig:OFFmore}, we show the distributions of the off-diagonal elements of the number operators of well 1 (a), 2 (b), and 3 (c) in comparison with Laplace, Logistic, and Gaussian distributions. Figure~\ref{Fig:OFFmore}~(a) is equivalent to Fig.~\ref{Fig:OFF270}~(a) in the main text. The best fit for the three observables is the Laplace distribution.

%


\end{document}